\input lanlmac
\def\href#1#2{{#2}}
\def\hhref#1{{#1}}
\input epsf.tex

\overfullrule=0mm

\newcount\figno
\figno=0
\def\fig#1#2#3{
\par\begingroup\parindent=0pt\leftskip=1cm\rightskip=1cm\parindent=0pt
\baselineskip=11pt
\global\advance\figno by 1
\midinsert
\epsfxsize=#3
\centerline{\epsfbox{#2}}
\vskip 12pt
{\bf Fig.\ \the\figno:} #1\par
\endinsert\endgroup\par
}
\def\figlabel#1{\xdef#1{\the\figno}}
\def\encadremath#1{\vbox{\hrule\hbox{\vrule\kern8pt\vbox{\kern8pt
\hbox{$\displaystyle #1$}\kern8pt}
\kern8pt\vrule}\hrule}}


\def\IR{\relax{\rm I\kern-.18em R}}
\font\cmss=cmss10 \font\cmsss=cmss10 at 7pt

\font\cmss=cmss10 \font\cmsss=cmss10 at 7pt
\def\IZ{\relax\ifmmode\mathchoice
{\hbox{\cmss Z\kern-.4em Z}}{\hbox{\cmss Z\kern-.4em Z}}
{\lower.9pt\hbox{\cmsss Z\kern-.4em Z}}
{\lower1.2pt\hbox{\cmsss Z\kern-.4em Z}}\else{\cmss Z\kern-.4em Z}\fi}
\def\IN{\relax{\rm I\kern-.18em N}}
\def\b{\circ}
\def\n{\bullet}

\def\gbbbb{\Gamma_4^{\hbox{$\scriptstyle \b \b$}\kern -8.2pt
\raise -4pt \hbox{$\scriptstyle \b \b$}}}
\def\gnnnn{\Gamma_4^{\hbox{$\scriptstyle \n \n$}\kern -8.2pt  
\raise -4pt \hbox{$\scriptstyle \n \n$}}}
\def\gnnnnnn{\Gamma_6^{\hbox{$\scriptstyle \n \n \n$}\kern -12.3pt
\raise -4pt \hbox{$\scriptstyle \n \n \n$}}}
\def\gbbbbbb{\Gamma_6^{\hbox{$\scriptstyle \b \b \b$}\kern -12.3pt
\raise -4pt \hbox{$\scriptstyle \b \b \b$}}}
\def\gbbbbc{\Gamma_{4\, c}^{\hbox{$\scriptstyle \b \b$}\kern -8.2pt
\raise -4pt \hbox{$\scriptstyle \b \b$}}}
\def\gnnnnc{\Gamma_{4\, c}^{\hbox{$\scriptstyle \n \n$}\kern -8.2pt
\raise -4pt \hbox{$\scriptstyle \n \n$}}}
\def\Rbud#1{{\cal R}_{#1}^{-\kern-1.5pt\blacktriangleright}}
\def\Rleaf#1{{\cal R}_{#1}^{-\kern-1.5pt\vartriangleright}}
\def\Rbudb#1{{\cal R}_{#1}^{\circ\kern-1.5pt-\kern-1.5pt\blacktriangleright}}
\def\Rleafb#1{{\cal R}_{#1}^{\circ\kern-1.5pt-\kern-1.5pt\vartriangleright}}
\def\Rbudn#1{{\cal R}_{#1}^{\bullet\kern-1.5pt-\kern-1.5pt\blacktriangleright}}
\def\Rleafn#1{{\cal R}_{#1}^{\bullet\kern-1.5pt-\kern-1.5pt\vartriangleright}}
\def\Wleaf#1{{\cal W}_{#1}^{-\kern-1.5pt\vartriangleright}}
\def\Cleaf{{\cal C}^{-\kern-1.5pt\vartriangleright}}
\def\Cbud{{\cal C}^{-\kern-1.5pt\blacktriangleright}}
\def\Crleaf{{\cal C}^{-\kern-1.5pt\circledR}}


\magnification=\magstep1
\baselineskip=12pt
\hsize=6.3truein
\vsize=8.7truein
 at 8truept
 at 8truept
 at 10truept

\font\bigrm=cmr12 at 14pt
\centerline{\bigrm Statistics of geodesics in large quadrangulations}

\bigskip\bigskip

\centerline{J. Bouttier and E. Guitter}
  \smallskip
  \centerline{Service de Physique Th\'eorique, CEA/DSM/SPhT}
  \centerline{Unit\'e de recherche associ\'ee au CNRS}
  \centerline{CEA/Saclay}
  \centerline{91191 Gif sur Yvette Cedex, France}
\centerline{\tt jeremie.bouttier@cea.fr}
\centerline{\tt emmanuel.guitter@cea.fr}

  \bigskip


     \bigskip\bigskip

     \centerline{\bf Abstract}
     \smallskip
     {\narrower\noindent
We study the statistical properties of geodesics, i.e.\ paths of minimal 
length, in large random planar quadrangulations. We extend Schaeffer's 
well-labeled tree bijection to the case of quadrangulations with a marked 
geodesic, leading to the notion of ``spine trees", amenable to a direct
enumeration. We obtain the generating functions for quadrangulations 
with a marked geodesic of fixed length, as well as with a set of 
``confluent geodesics", i.e.\ a collection of non-intersecting minimal paths 
connecting two given points. In the limit of quadrangulations with a large 
area $n$, we find in particular an average number $3 \times 2^i$ of 
geodesics between two fixed points at distance $i \gg 1$ from each other. 
We show that, for generic endpoints, two confluent 
geodesics remain close to each other and have an extensive
number of contacts. This property fails for a few ``exceptional" endpoints 
which can be linked by truly distinct geodesics. Results are presented both 
in the case of finite length $i$ and in the scaling limit 
$i\propto n^{1/4}$. In particular, we give the scaling distribution of
the exceptional points.
\par}

     \bigskip

\nref\QGRA{V. Kazakov, {\it Bilocal regularization of models of random
surfaces}, Phys. Lett. {\bf B150} (1985) 282-284; F. David, {\it Planar
diagrams, two-dimensional lattice gravity and surface models},
Nucl. Phys. {\bf B257} (1985) 45-58; J. Ambjorn, B. Durhuus and J. Fr\"ohlich,
{\it Diseases of triangulated random surface models and possible cures},
Nucl. Phys. {\bf B257}(1985) 433-449; V. Kazakov, I. Kostov and A. Migdal
{\it Critical properties of randomly triangulated planar random surfaces},
Phys. Lett. {\bf B157} (1985) 295-300.}
\nref\TUT{W. Tutte,
{\it A Census of planar triangulations} Canad. J. of Math. {\bf 14} (1962) 21-38;
{\it A Census of Hamiltonian polygons} Canad. J. of Math. {\bf 14} (1962) 402-417;
{\it A Census of slicings}, Canad. J. of Math. {\bf 14} (1962) 708-722;
{\it A Census of Planar Maps}, Canad. J. of Math. {\bf 15} (1963) 249-271.
}
\nref\BIPZ{E. Br\'ezin, C. Itzykson, G. Parisi and J.-B. Zuber, {\it Planar
Diagrams}, Comm. Math. Phys. {\bf 59} (1978) 35-51.}
\nref\DGZ{P. Di Francesco, P. Ginsparg and J. Zinn--Justin, {\it 2D Gravity and Random Matrices},
Physics Reports {\bf 254} (1995) 1-131.}
\nref\AW{J. Ambj\o rn and Y. Watabiki, {\it Scaling in quantum gravity},
Nucl.Phys. {\bf B445} (1995) 129-144.}
\nref\AJW{J. Ambj\o rn, J. Jurkiewicz and Y. Watabiki,
{\it On the fractal structure of two-dimensional quantum gravity},
Nucl.Phys. {\bf B454} (1995) 313-342.}
\nref\MS{M. Marcus and G. Schaeffer, {\it Une bijection simple pour les
cartes orientables}, available at 
\hhref{http://www.lix.polytechnique.fr/Labo/Gilles.Schaeffer/Biblio/}.}
\nref\CS{P. Chassaing and G. Schaeffer, {\it Random Planar Lattices and 
Integrated SuperBrownian Excursion}, 
Probability Theory and Related Fields {\bf 128(2)} (2004) 161-212, 
arXiv:math.CO/0205226.}
\nref\MOB{J. Bouttier, P. Di Francesco and E. Guitter. {\it 
Planar maps as labeled mobiles},
Elec. Jour. of Combinatorics {\bf 11} (2004) R69, arXiv:math.CO/0405099.}
\nref\FOMAP{J. Bouttier, P. Di Francesco and E. Guitter. {\it Blocked edges 
on Eulerian maps and mobiles: Application to spanning trees, hard particles 
and the Ising model}, 	J. Phys. A: Math. Theor. {\bf 40} (2007) 7411-7440, 
arXiv:math.CO/0702097.}
\nref\GEOD{J. Bouttier, P. Di Francesco and E. Guitter, {\it Geodesic
distance in planar graphs}, Nucl. Phys. {\bf B663}[FS] (2003) 535-567, 
arXiv:cond-mat/0303272.}
\nref\MARMO{J. F. Marckert and A. Mokkadem, {\it Limit of normalized
quadrangulations: the Brownian map}, arXiv:math.PR/0403398.}
\nref\LEGALL{J. F. Le Gall, {\it The topological structure of scaling limits 
of large planar maps}, arXiv:math.PR/0607567.}
\nref\LGP{J. F. Le Gall and F. Paulin,
{\it Scaling limits of bipartite planar maps are homeomorphic to the 2-sphere},
arXiv:math.PR/0612315.}
\nref\CORV{R. Cori and B. Vauquelin, {\it Planar maps are well labeled trees},
Canad. J. Math. {\bf 33 (5)} (1981) 1023-1042.}

\newsec{Introduction}

The study of random maps is a fundamental issue both in mathematics, where
it raises many interesting combinatorial and probabilistic problems, and in 
physics, where maps serve as discretizations for fluctuating surfaces in
various domains. Understanding the statistical properties of ensembles of 
random maps is relevant in particular in the context of two-dimensional 
quantum gravity \QGRA\ or for the description of fluid membrane statistics. 
So far, most results deal with global properties of random maps, with 
much effort made on the {\it enumeration} of families of maps. Various 
methods were developed in this context, such as Tutte's original approach 
through recursive decomposition \TUT\ or the more technical framework of 
random matrix integrals [\xref\BIPZ,\xref\DGZ]. 
Correlations functions of various observables were obtained for ensembles 
of maps with possibly extra statistical degrees of freedom such as spins or 
particles, but mostly in a global way consisting on averaging over the
positions of the points where these observables are measured (see \DGZ). 
Much less is known on more refined local properties of maps, such as the 
actual dependence of correlations on the distance between the points 
where the observables are measured [\xref\AW,\xref\AJW].
In this respect, a particularly promising approach relies on a recent 
{\it bijective} enumeration technique which consists in a coding of maps by 
decorated trees. This approach was initiated by Schaeffer who found a 
one-to-one mapping between, on the one hand, planar quadrangulations, i.e.\ maps
with tetravalent faces only, and on the other hand so-called 
{\it well-labeled trees}, i.e.\  plane trees with suitably 
labeled vertices [\xref\MS,\xref\CS]. 
His construction was then extended to various families of 
maps, such as maps with prescribed faces valences \MOB, Eulerian
maps and even maps with particles or spins \FOMAP. The main advantage of the 
approach is that it keeps track of the graph distance of vertices from 
a given arbitrary origin vertex, thus giving access to statistical properties
of maps controlled by this distance. A first application concerned 
the so-called {\it two-point function}, i.e.\ the average number of 
pairs of vertices at a fixed distance from each other among quadrangulations 
of fixed area. In particular, for large maps, this two-point function was 
shown to converge to a universal scaling function characterizing  
the density of vertices at a fixed distance from an arbitrary origin
in the continuous scaling limit of large maps [\xref\AW,\xref\GEOD]. 

In this paper, we address the question of the statistics of {\it geodesics},
i.e.\ paths which join two given vertices in a random quadrangulation, and 
are of minimal length among all paths linking these vertices. Here again, 
we rely on an extension of Schaeffer's bijection using well-labeled trees
to code for quadrangulations with a marked geodesic path. In this
setting, we have access to the statistical properties of geodesics 
with a fixed given length, as well as of {\it confluent geodesics}, 
i.e.\ collections of non-intersecting minimal paths connecting the same two
vertices. Many results can be explicited in the limit of large quadrangulations,
i.e.\ when the area of the quadrangulation tends to infinity.
Two different regimes may be studied according to whether the length
of the geodesics remains finite or scales as the appropriate power
of the diverging area. This latter scaling regime is particularly
important to extract universal properties characterizing the {\it continuous 
limit of large quadrangulations}, a probabilistic object whose construction
is still under investigation [\xref\MARMO-\xref\LGP].

The paper is organized as follows: Section 2 is devoted to the combinatorics
of quadrangulations with marked geodesics. We first recall in
Section 2.1  Schaeffer's construction which associates to each quadrangulation 
a well-labeled tree. We then show in Section 2.2 how to extend this bijection
to treat the case of quadrangulations with a marked geodesic path. 
This leads us to the notion of {\it spine tree}, i.e.\ a well-labeled
tree with a distinguished linear subtree (the spine) inherited from the
marked path. Spine trees are shown to be in bijection with quadrangulations
equipped with a boundary made of two geodesics. Under the condition
that these two boundary geodesics have no intermediate contact, the configurations
can be continuously transformed into the desired configurations of 
quadrangulations with a single marked geodesic path. In Section 2.3, we 
derive explicit formulae for the generating function of 
quadrangulations bounded by two geodesic paths and for that of quadrangulations
with a single marked geodesic of given length. The connection between the 
two generating functions is obtained via a simple transformation consisting
in taking an irreducible part. 
We finally show in Section 2.4 how to use our basic generating functions 
to enumerate quadrangulations with marked confluent geodesics. We distinguish
two general cases: weakly avoiding confluent geodesics made of paths that
can touch each other with no intersections or strongly avoiding confluent
geodesics which have no contact vertices except for their common first and 
last vertices.
Section 3 deals with the statistical properties of geodesics of 
{\it finite length} $i$ in large quadrangulations, i.e.\ in an ensemble of 
quadrangulations with fixed area $n$ in the limit $n\to \infty$, keeping
$i$ finite. We first explain in Section 3.1 how our enumerative results
can be translated into averages in the ensemble of pointed quadrangulations
of fixed area $n$ (with inverse symmetry weights) or the ensemble of 
quadrangulations with two marked points at fixed distance.
We then give in Section 3.2 the generating function for the average number 
of geodesics of length $i$ in pointed quadrangulations with area 
$n\to \infty$. In particular,
for large $i$, we obtain an average number $3\times 2^i$ of 
geodesics linking two given vertices at distance $i$. The average number
of confluent geodesics is computed in Section 3.3 and some of their
refined properties are discussed in Section 3.4, such as the
average number of contacts between two weakly avoiding geodesics and 
the average area which they enclose. We finally discuss the case of
strongly avoiding geodesics and show that, at large $n$, only a set
of size $\propto n^{1/4}$ of {\it exceptional} points can be reached from a given
vertex by two such strongly avoiding geodesics. 
In Section 4, we study a different scaling regime in which the
size $n$ of the quadrangulation and the length $i$ of the geodesics
tend simultaneously to infinity, keeping the ratio $r=i/n^{1/4}$ fixed. This scaling
is customary in the context of large quadrangulations and we recall in
Section 4.1 the formula for the universal density $\rho(r)$ of vertices 
at rescaled distance $r$ from a given origin. We show in Section 4.2 that
upon appropriate renormalizations, the average number of geodesics 
of rescaled length $r$, or of $k$-tuples of confluent geodesics is 
characterized by the same scaling function $\rho(r)$. On the other hand, we
show that the density of exceptional points is described by a new
scaling function that we compute in Section 4.3. We finally discuss in
Section 4.4 the area enclosed by 
two geodesics and show that this area is proportional to the total area
$n$ only in the case of strongly avoiding geodesics. We discuss our
results and conclude in Section 5. 

\newsec{Geodesics in quadrangulations and labeled trees}

\subsec{Schaeffer's construction for planar quadrangulations}

\fig{The two possible types of faces in a planar quadrangulation according
to the distances of their incident vertices from a given origin: (a)
simple faces and (b) confluent faces. To each face, we associate an edge
of well-labeled tree (thick solid or dashed line).}{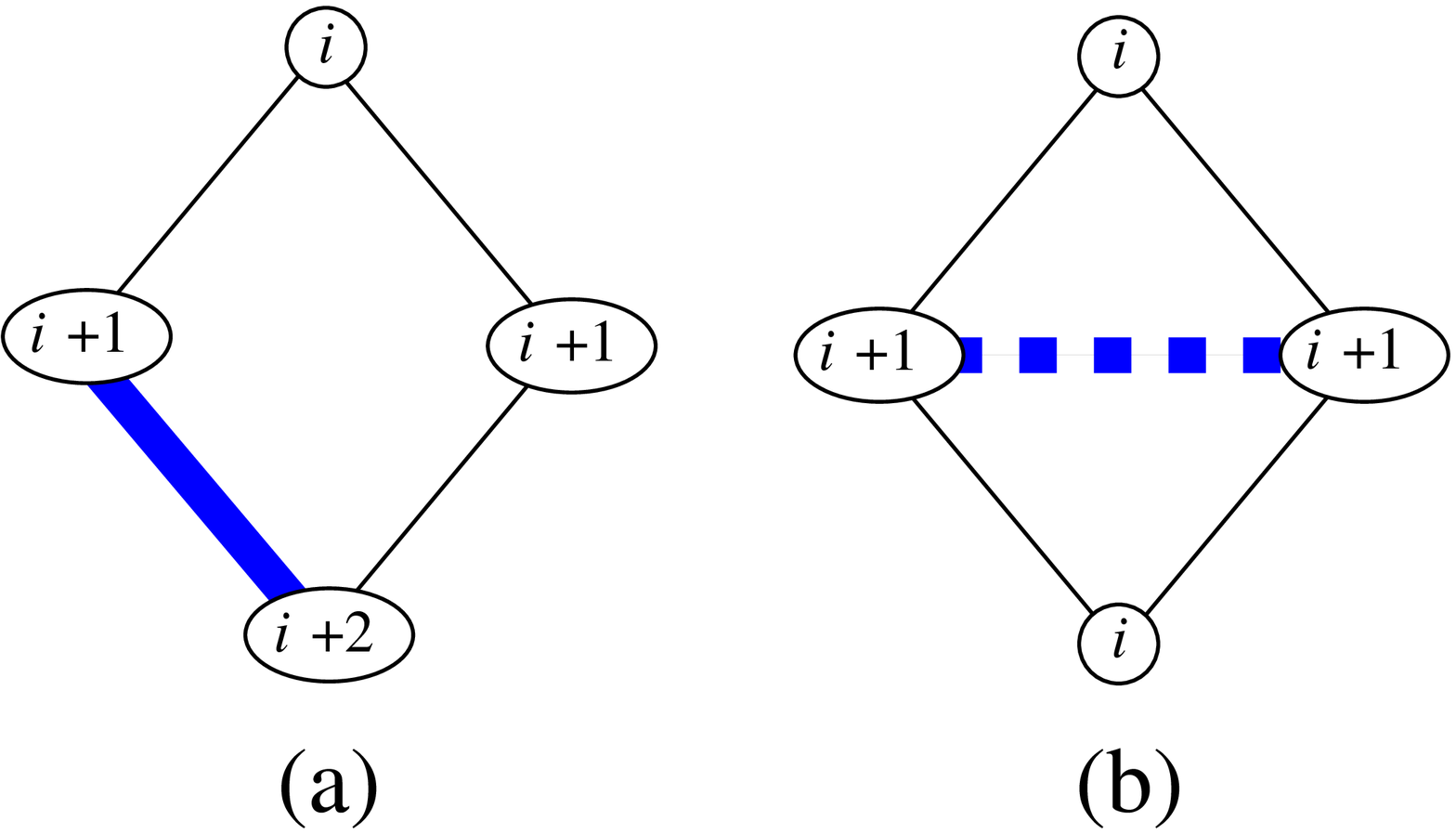}{8.cm}
\figlabel\faces
In order to study maps with marked geodesics, we shall recourse to the 
well-known correspondence between rooted planar quadrangulations and 
so-called well-labeled trees \CORV\ in the form found by Schaeffer
{\it et al.} [\xref\MS,\xref\CS]. The correspondence 
works as follows: starting from a rooted planar quadrangulation,
i.e.\ a quadrangulation of the sphere with a distinguished oriented edge, 
we label each vertex by its graph distance from the {\it origin}, 
i.e.\ the vertex from which the root edge originates. Each vertex therefore
receives a strictly positive integer label, except for the origin which
receives the label $0$. It is easily seen that the faces of the 
quadrangulation fall into two classes (see Fig.\faces): simple faces where
the cyclic sequence of labels around the face is of the form $i$, $i+1$,
$i+2$, $i+1$ for some $i\geq 0$ and confluent faces where it is
of the form $i$, $i+1$, $i$, $i+1$. 
\fig{An example (a) of rooted quadrangulation (thick or thin solid lines) 
and the associated edges (thick solid or dashed lines) in Schaeffer's 
construction. These edges form a planted well-labeled 
tree (b).}{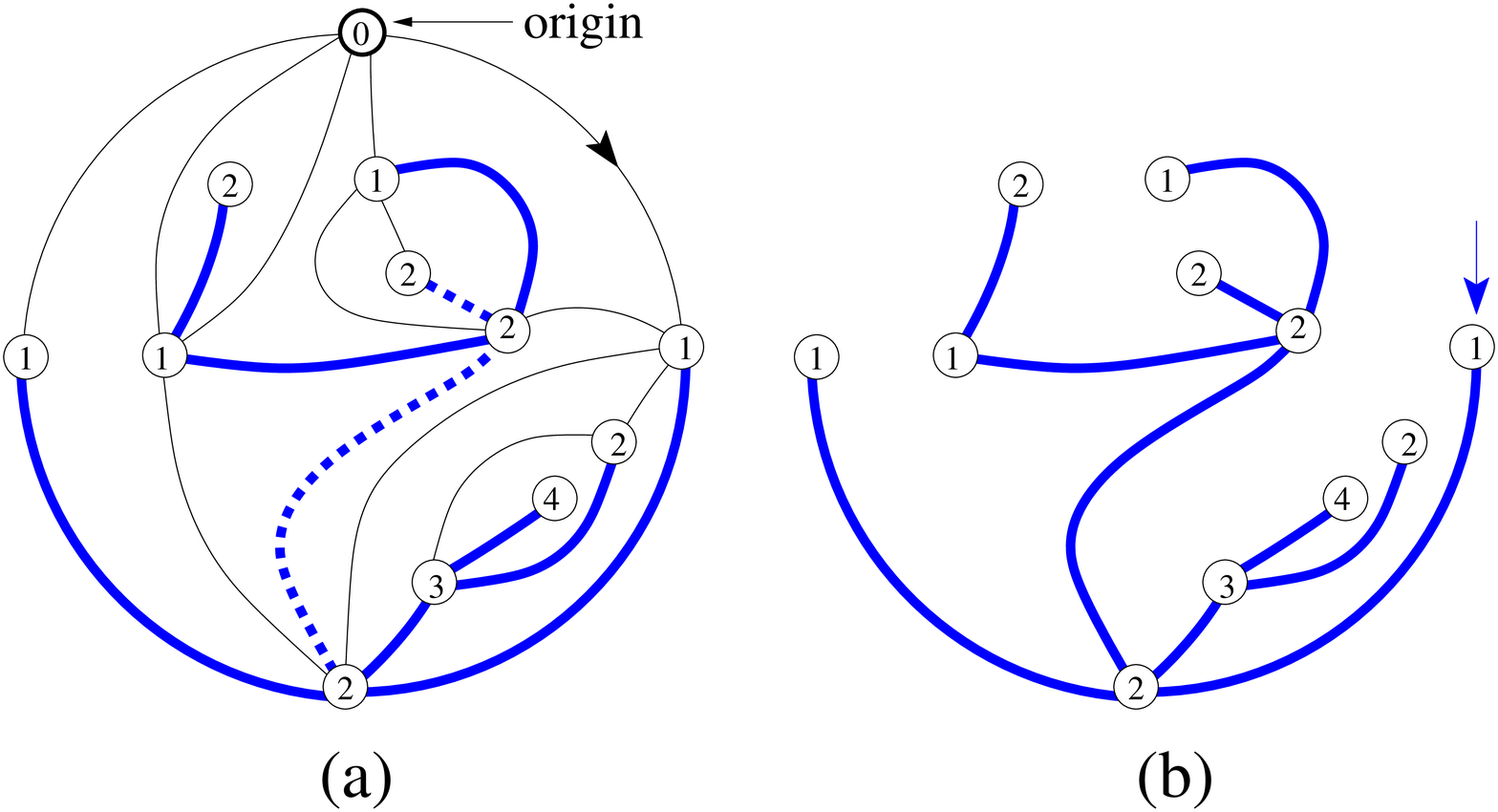}{12.cm}
\figlabel\quadtotree
For each simple face, we select the unique edge of type $i+2 \to i+1$ 
encountered clockwise around the face (see Fig.\faces-(a)). 
For each confluent face, we draw a new edge 
linking the two vertices labeled $i+1$ (see Fig.\faces-(b)). The procedure thus
associates an edge to each face of the original quadrangulation and
it can be shown that the graph made of these edges connects all the original 
vertices except the origin that remains isolated. Furthermore, it is
planar, contains no loop and has one more vertex than edge, hence it is 
a plane tree (see Fig.\quadtotree\ for
an example). By construction, the labels on the vertices are strictly 
positive integers and vary by at most one along any edge of the tree. 
This property characterizes so-called {\it well-labeled} trees. The root 
edge of the quadrangulation selects a corner of the tree at its endpoint 
(labeled $1$), at which we plant the tree. 
It was shown that the construction above provides a bijection between rooted 
planar quadrangulations and well-labeled trees planted at a corner labeled $1$.
\fig{Construction of the quadrangulation associated with a given well-labeled
tree. Corners are connected to their successors (a) by a set of
non-crossing arches (dashed lines). These arches form the edges of the
quadrangulation (b).}{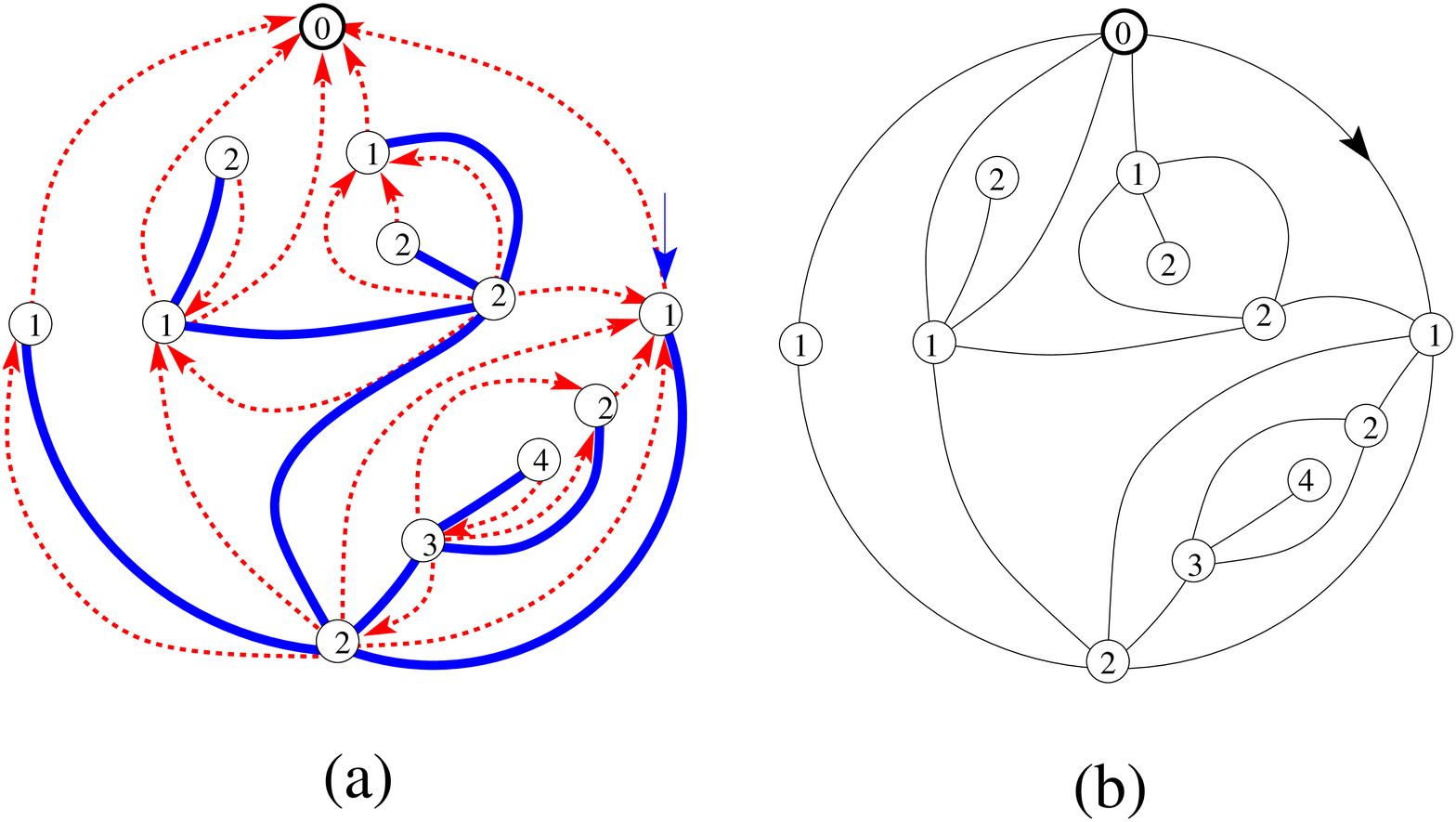}{12.cm}
\figlabel\treetoquad
The inverse construction can be realized as follows: starting from
a well-labeled tree, we connect by an arch each corner with label $i>1$ to its
{\it successor}, that is the first corner labeled $i-1$ encountered clockwise
around the tree (see Fig.\treetoquad\ for an example). 
Corners labeled $1$ are then connected by arches to an
extra vertex $0$ outside the tree. The fact that labels vary by at most one 
along an edge ensures that the arches can be drawn with no intersection, 
resulting in a planar quadrangulation while the labels simply
record the graph distance from the added vertex. The arch from the corner
labeled $1$ at which the tree was planted to the extra vertex, when
oriented backwards, defines the root edge of the quadrangulation.
\eject
\subsec{Extension to quadrangulations with a marked geodesic}
\fig{The extension of Schaeffer's construction to quadrangulations
with a marked geodesic. In (a), we only display the marked geodesic, while
the underlying quadrangulation is symbolized by the grey background. 
We first unzip the geodesic path (b) and fill the created hole by 
squares (c). Applying the rules of Fig.\faces\ for the added squares
selects all the edges of the left copy of the geodesic but one (d),
creating a spine (thick line). The original faces of the quadrangulation
give rise to well-labeled subtrees attached only to 
the left side of this spine, 
resulting in a spine tree (e).}{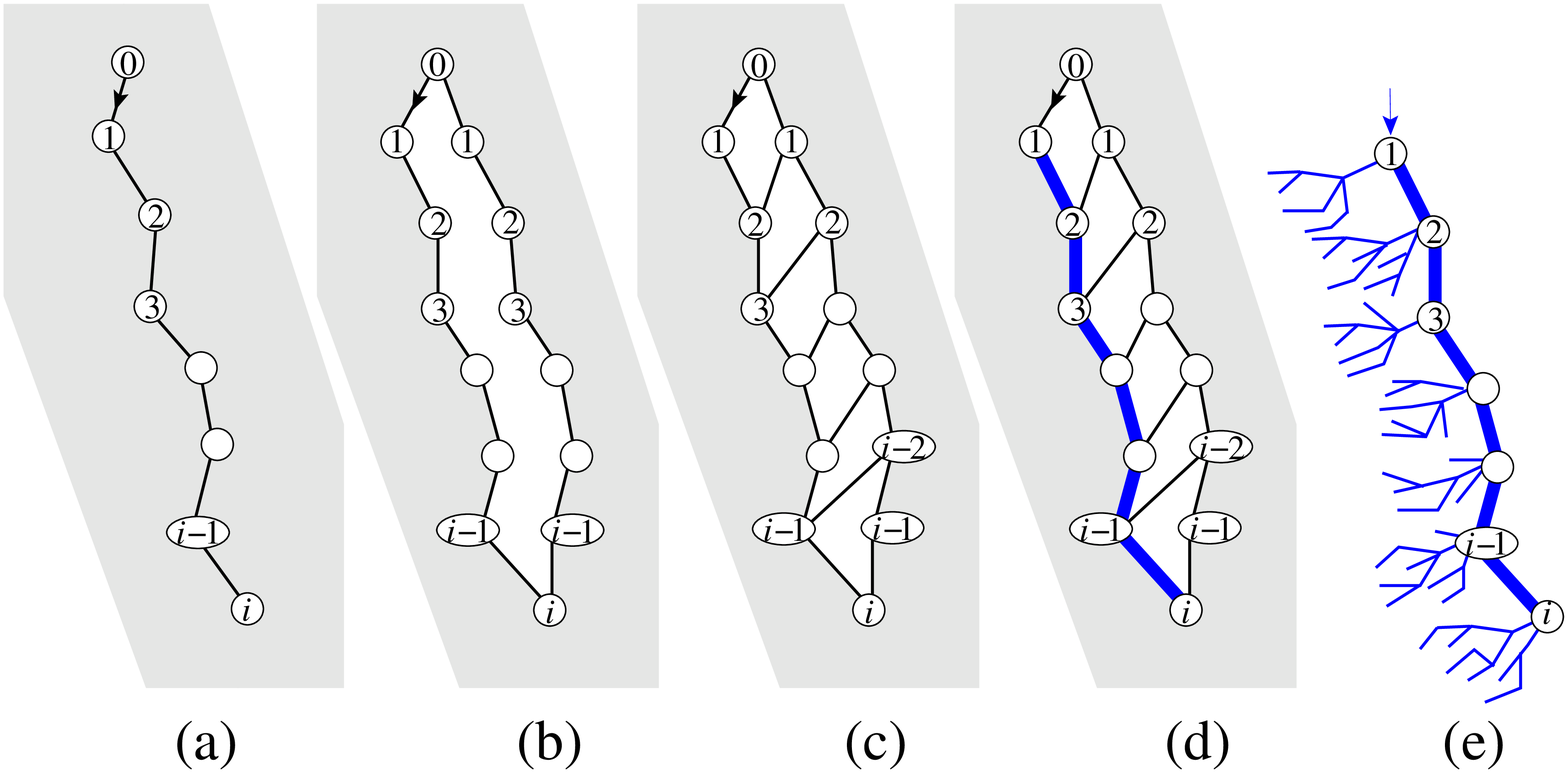}{13.cm}
\figlabel\cutgeod
We now turn to the study of quadrangulations with a marked geodesic,
that is an oriented path connecting two vertices and having minimal length 
among all paths connecting these vertices. The vertex from which the geodesic
originates will be referred to as the origin vertex. Clearly, when moving 
along a geodesic, the distance from the origin of the successive vertices 
increases by one at each step. Note
that picking a geodesic of length $1$ simply amounts to marking an oriented 
edge of the quadrangulation, in which case the object at hand is nothing but 
a rooted quadrangulation on which we may apply Schaeffer's construction. 
If the length $i$ of the geodesic is larger or equal to $2$, it is then 
convenient to adapt this construction to ensure that all the edges of the 
geodesic path but the first one are selected and form a path of length $i-1$ 
on the tree. To this end, we decide to first ``unzip" the geodesic so 
as to form a hole in the quadrangulation, bounded by two copies of the original 
geodesic (see Fig.\cutgeod), which we will refer to as the left and right
copy upon viewing the hole with the origin at the top. We then fill up
the hole with $i-1$ new squares obtained by adding $i-2$ parallel edges 
linking, for each $k=1, \cdots, i-2$, the vertex of the right boundary at
distance $k$ from the origin to the vertex of the left boundary at 
distance $k+1$. At this stage, it is important to note that both the 
unzipping process and the filling process {\it do not affect the distances}
from the origin on the quadrangulation. In particular, both copies of
the original geodesic are actual geodesic paths in the modified quadrangulation
and their vertex labels indeed correspond to the distance from
the origin. We finally root the modified quadrangulation 
by selecting the edge $0 \to 1$ of the left copy of the geodesic. 

With this rooted quadrangulation at hand, we may now apply Schaeffer's
construction and transform it into a well-labeled tree. By construction,
all the added squares are simple faces of which we select the lower most 
edge on the left boundary (see Fig.\cutgeod-(d)). All the edges of the 
left boundary but the first one are therefore selected, thus forming a path of 
length $i-1$ on the tree linking $i$ vertices with increasing labels 
$1,2,\cdots,i$. We shall call this path the spine of the tree. On the other 
hand, none of the added edges is selected and the rest of the tree is 
therefore made of $i$ (possibly empty) subtrees attached only 
{\it to the left side} of the spine, when viewing it as hanging from its 
first vertex. 

In view of this result, we are naturally led to define a {\it spine tree}
as a particular type of well-labeled tree, made of a spine of
length $i-1$ joining $i$ vertices with increasing labels $1,2,\cdots,i$ and
of $i$ well-labeled subtrees attached to the left side of this spine.
The tree is planted at the first vertex of the spine and can be viewed as 
hanging from that vertex. 

\fig{Starting from a spine tree obtained as in Fig.4, we reconstruct 
the two copies of the geodesic
path by linking corners to their successors (a). Corners on the left
of the spine create the first copy (green short-dashed arches), while the 
chain of 
successors of the vertex $i$ at the end of the spine creates the other copy
(red long-dashed arches), the two copies being connected by winding
(magenta dot-dashed) arches. 
They moreover surround all subtrees hanging from the 
spine hence, after completing the inverse construction, serve as 
boundaries for the rest of the quadrangulation (grey background in (b)).
On the sphere, this configuration
can be deformed into a quadrangulation with an internal zipper-like structure 
(c) which can be eventually squeezed into a marked geodesic 
(d).}{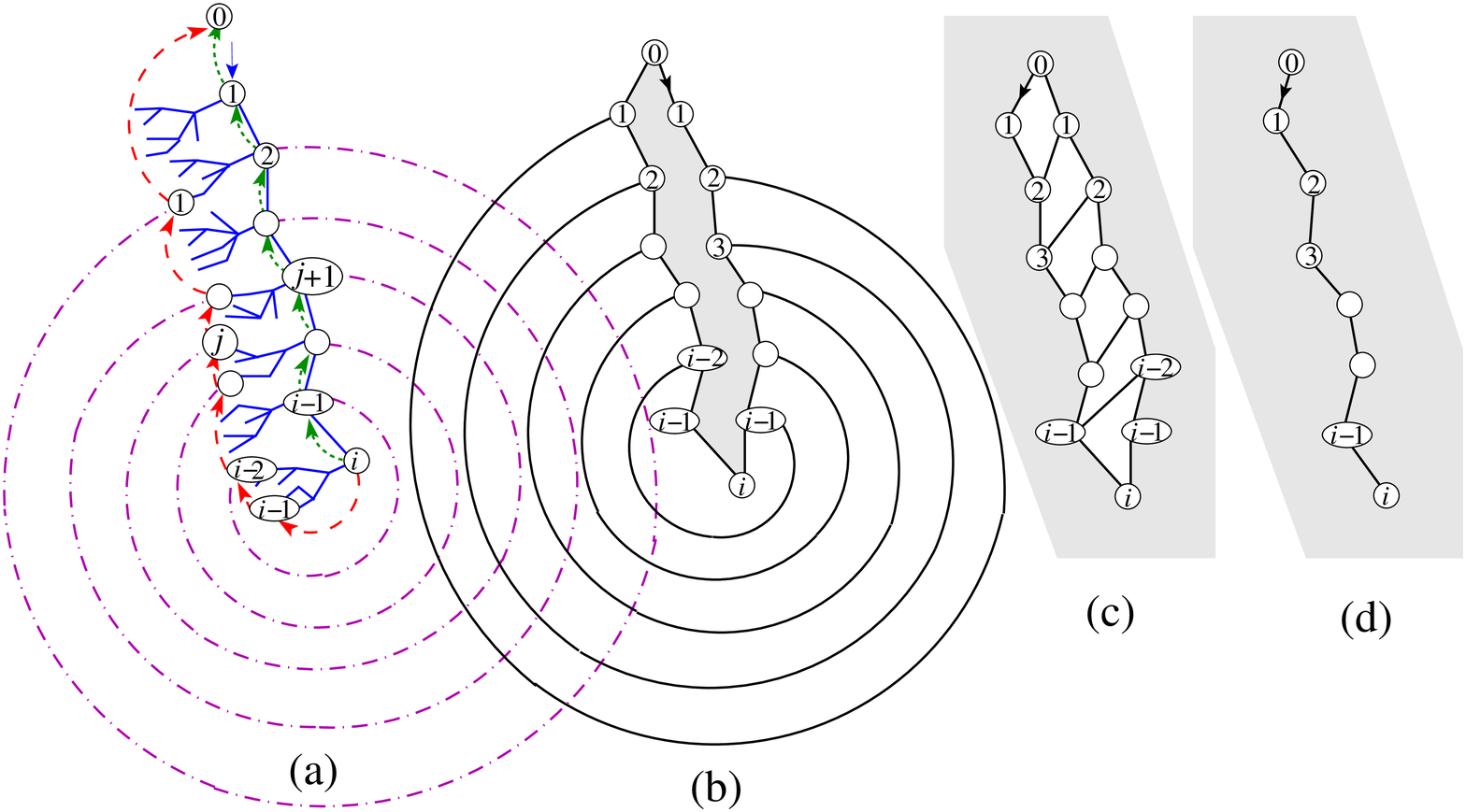}{14.cm}
\figlabel\backtogeod
With this definition, we see that applying Schaeffer's construction to our
modified quadrangulation results in a spine tree. As we shall see below,
not all spine trees however are obtained that way. This can be understood
by first applying the inverse construction recalled in Sect.2.1 to a spine
tree obtained as above from a quadrangulation with a marked geodesic 
and following
how the two copies of the geodesic path are reconstructed. 
For each vertex along the spine with label $\geq 2$, 
the associated corner lying just on the left 
of the incident spine edge pointing toward the root gives rise 
to an arch linking that corner to the preceding vertex on the spine 
(see Fig.\backtogeod). This set of arches, together with the arch linking 
the root corner to the extra origin vertex labeled $0$, forms the left copy 
of the geodesic path on the modified quadrangulation. On the other hand, 
consider now the vertex labeled $i$ at the end of the spine and its chain
of successors  (i.e.\ its successor, the successor of its successor, etc):
this chain forms the right copy of the geodesic. Furthermore, one easily 
sees that, for $j\geq 1$, the corner labeled $j+1$ lying on the right of 
the spine will be connected by 
an arch winding around the tree to the vertex labeled $j$ of the chain. 
Completing the inverse construction, 
the resulting object is the desired quadrangulation where the two copies of the 
geodesic are now connected by $i-2$ winding arches, with the rest of the 
quadrangulation in between. Upon continuous deformation on the sphere, we  
can recover the wanted zipper-like structure with the former winding arches now lying in 
between the two paths (see Fig.\backtogeod). This last structure is eventually 
zipped so as 
to recover the marked geodesic on the quadrangulation. 
\fig{Appearance of pinch points in the inverse construction from a 
generic spine tree. If the successor of a corner labeled $j+1$ on the
right of the spine happens to be itself on the spine (a), this creates a
pinch point at distance $j$ from the origin in the associated 
quadrangulation (b). Here we have not represented the winding arches,
as opposed to Fig.\backtogeod.} {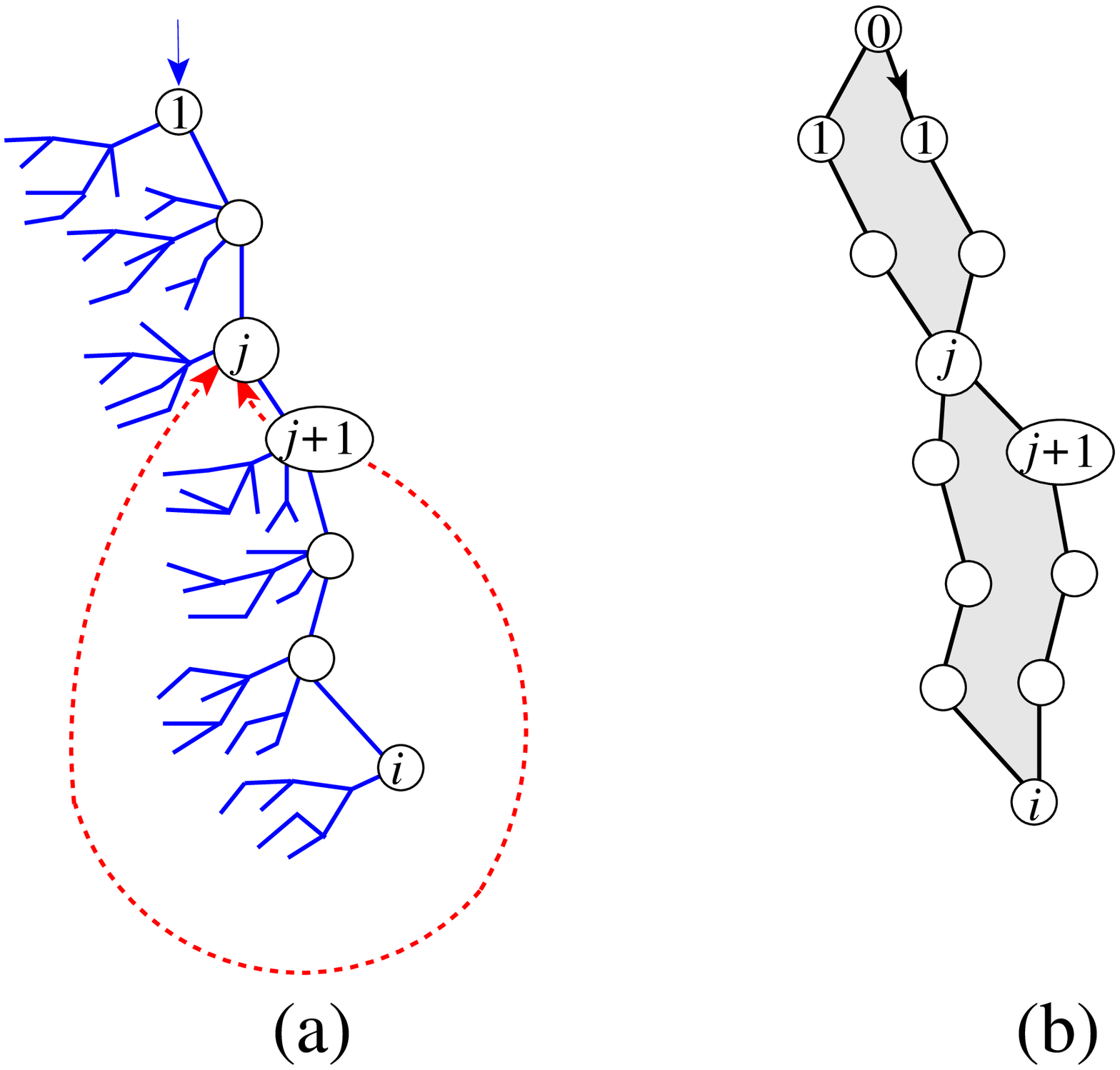}{8.cm}
\figlabel\coincidence
Now we see that, in the last transformation, it is crucial that the two 
copies of the geodesic path obtained by the inverse construction have no 
common vertex (except for their first and last vertex). This is the case
for a spine tree resulting from the modified Schaeffer's construction
on a quadrangulation with a marked geodesic, but this is not true in 
general for an arbitrary spine tree. Indeed, for an arbitrary spine tree,
it may happen that the successor 
of a corner on the right of the spine lies itself on the spine 
(see Fig.\coincidence), necessarily on its left side. If so, we will obtain 
two geodesic paths with a common vertex acting as a {\it pinch point}
separating the 
rest of the quadrangulation into two parts (see Fig.\coincidence).
\fig{A schematic picture (a) of a quadrangulation with a geodesic boundary
of length $2i$, as obtained from the inverse construction acting on
a generic spine tree. The two geodesic paths bounding the quadrangulation
may have pinch points in the form of common vertices and/or common edges.
Quadrangulations with a geodesic boundary of length $2i$ having no pinch point
(b) are in bijection with quadrangulations with a marked geodesic of
length $i$.}{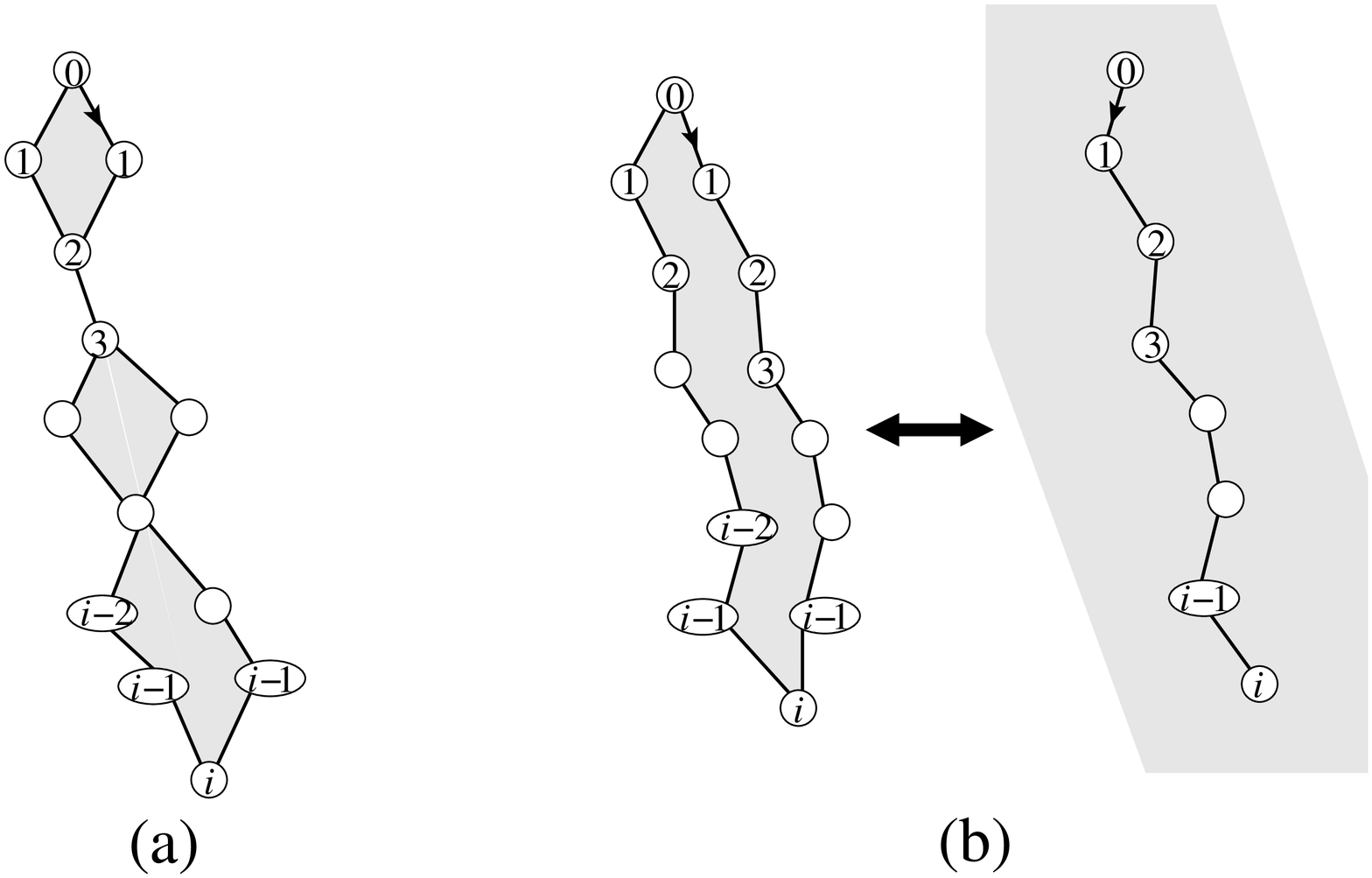}{10.cm}
\figlabel\bord
When applying the inverse construction to a general well-labeled spine tree 
with spine of length $i-1$, if we decide not to draw the winding arches, 
the most general object that we end up with is 
a {\it quadrangulation with a geodesic boundary} of length $2i$, 
i.e.\ a boundary with a marked origin vertex and such that the two
paths of length $i$ joining this origin to its antipodal vertex are geodesic
paths (see Fig.\bord\ for an illustration). 
This requirement is equivalent to demanding that the vertex antipodal 
to the origin be indeed at distance $i$ from this origin in the whole 
quadrangulation, i.e.\ all paths in the bulk of
the quadrangulation from the origin to that vertex have length larger or 
equal to $i$. The two geodesic paths forming the boundary may however meet 
at common vertices (with necessarily the same label on both sides) or even 
along common edges. The construction above is clearly one-to-one.

This constitutes {\it bijection I} between, on the one hand, 
quadrangulations with a geodesic boundary of length $2i$ 
(and possibly pinch points) and, on the other hand, planted 
well-labeled spine trees with spine of length $i-1$. 
By convention, for $i=1$, a well-labeled spine tree with spine of 
length $0$ is nothing but a well-labeled tree planted at a corner labeled $1$.

Moreover, there is a straightforward {\it bijection II} between 
quadrangulations with a geodesic boundary of length $2i$ {\it having no 
pinch point} and quadrangulations with a marked geodesic of length $i$, as
obtained by gluing the two sides of the boundary around the sphere 
(see Fig.\bord). 

We could easily characterize the subset of spine trees which are in
correspondence with geodesic boundaries without pinch points. 
It is however much simpler to note that a quadrangulation whose boundary 
has $k$ pinch points and length $2i$ may be uniquely decomposed into $k+1$ 
{\it irreducible components}
whose boundary lengths add up to $2i$. Each of these irreducible 
components is either a single edge or is precisely a quadrangulation with a 
geodesic boundary having no pinch point. This constitutes {\it bijection III}
and is the final step that will allow us to relate the enumeration
of quadrangulations with a marked geodesic to that of spine trees. 

\subsec{Generating functions}
We now would like to count quadrangulations with marked geodesics.
As usual, we will consider the generating function of such maps with
a weight $g$ per face. In the well-labeled tree language, this translates
into a weight $g$ per edge of the tree. We shall call $R_i(g)$ the 
generating function for well-labeled trees planted at
a corner labeled $i$. From Schaeffer's bijection, $R_1(g)$ is nothing
but the generating function of rooted planar quadrangulations or planar
quadrangulations with a marked geodesic of length $1$. 
As mentioned in Ref.\MOB, it can be shown that the quantity 
$\log\big(R_i(g)/R_{i-1}(g)\big)$ is the generating function for
quadrangulations with two marked (and distinguished) vertices at 
distance $i$ from each other (counted with symmetry factors). 
The generating
functions $R_i(g)$ satisfy the recursion relations
\eqn\relR{R_i(g)={1\over 1-g (R_{i-1}(g)+R_i(g)+R_{i+1}(g))}}
for $i\geq 1$ with the convention $R_0=0$. This relation simply states
that a well-labeled tree planted at a vertex labeled $i$ is formed by 
a collection of an arbitrary number of subtrees hanging from that vertex which
are themselves well-labeled trees with a root vertex labeled $i-1$, $i$ or
$i+1$. These recursion relations were solved in Ref.\GEOD, with the result:
\eqn\resR{\eqalign{& R_i(g) =R(g) {(1-x(g)^i)(1-x(g)^{i+3})\over (1-x(g)^{i+1})
(1-x(g)^{i+2})}\cr & \hbox{where}\ \ R(g) = {1-\sqrt{1-12 g}\over 6g} \cr 
& \hbox{and}\ \ x(g) =
{1-24 g-\sqrt{1-12 g}+\sqrt{6} \sqrt{72 g^2+6
   g+\sqrt{1-12 g}-1} \over 2 \left(6 g+\sqrt{1-12
   g}-1\right)}\ .\cr}}
In this expression, the functions $R(g)$ and $x(g)$ are explicit 
solutions of the equations:
\eqn\eqRx{\eqalign { & R(g)= 1+3 g \, \big(R(g)\big)^2 \cr &
x(g)+{1\over x(g)} +1 ={1\over g \big(R(g)\big)^2}\ .\cr}}
All functions above have a singularity at $g=g_{\rm crit}=1/12$, which
governs the asymptotic behavior of the general term in their series
expansions in $g$. 
{}From $R_i(g)$, it is easy to obtain the generating function 
$Z_i(g)$ for well-labeled spine trees with a spine of length $i-1$ as 
these trees are simply made of a spine with $i$ attached well-labeled 
subtrees whose root 
vertices have labels increasing from $1$ to $i$. We immediately get:
\eqn\resZ{Z_i(g)=\prod_{j=1}^i R_j(g) = R(g)^i {(1-x(g))(1-x(g)^{i+3})
\over (1-x(g)^3) (1-x(g)^{i+1})}\ .}
For convenience, we decided not to assign the weight $g$ to the
edges of the spine. Indeed, in our construction, these edges 
correspond to the added squares which were not present in the
original quadrangulation, and should therefore not be counted
as real squares. From bijection I of previous section, $Z_i(g)$ is
nothing but the generating function of quadrangulations with a 
geodesic boundary of length $2i$, as defined above. 
\fig{The set of all quadrangulations with a geodesic boundary of 
length $4$ and with $0$ or $1$ face (in grey).}{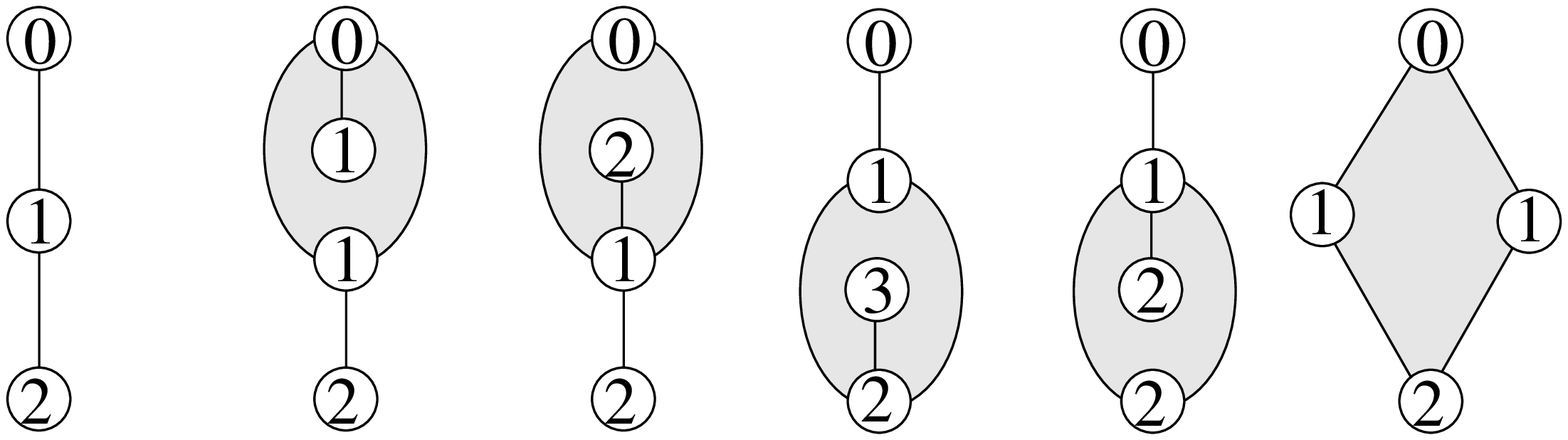}{10.cm}
\figlabel\zdeux
\noindent For instance,
we have the expansion $Z_2(g)=1+5 g+ 32 g^2 +234 g^3 +\ldots$ where
the first two terms correspond to the configurations displayed in Fig.\zdeux .

\fig{A schematic picture of the configurations of quadrangulations with a 
geodesic boundary of length $2i$ as counted by $Z_i$ (generic case) 
or $U_i$ (case
with no pinch point), and of their relation (2.5).}{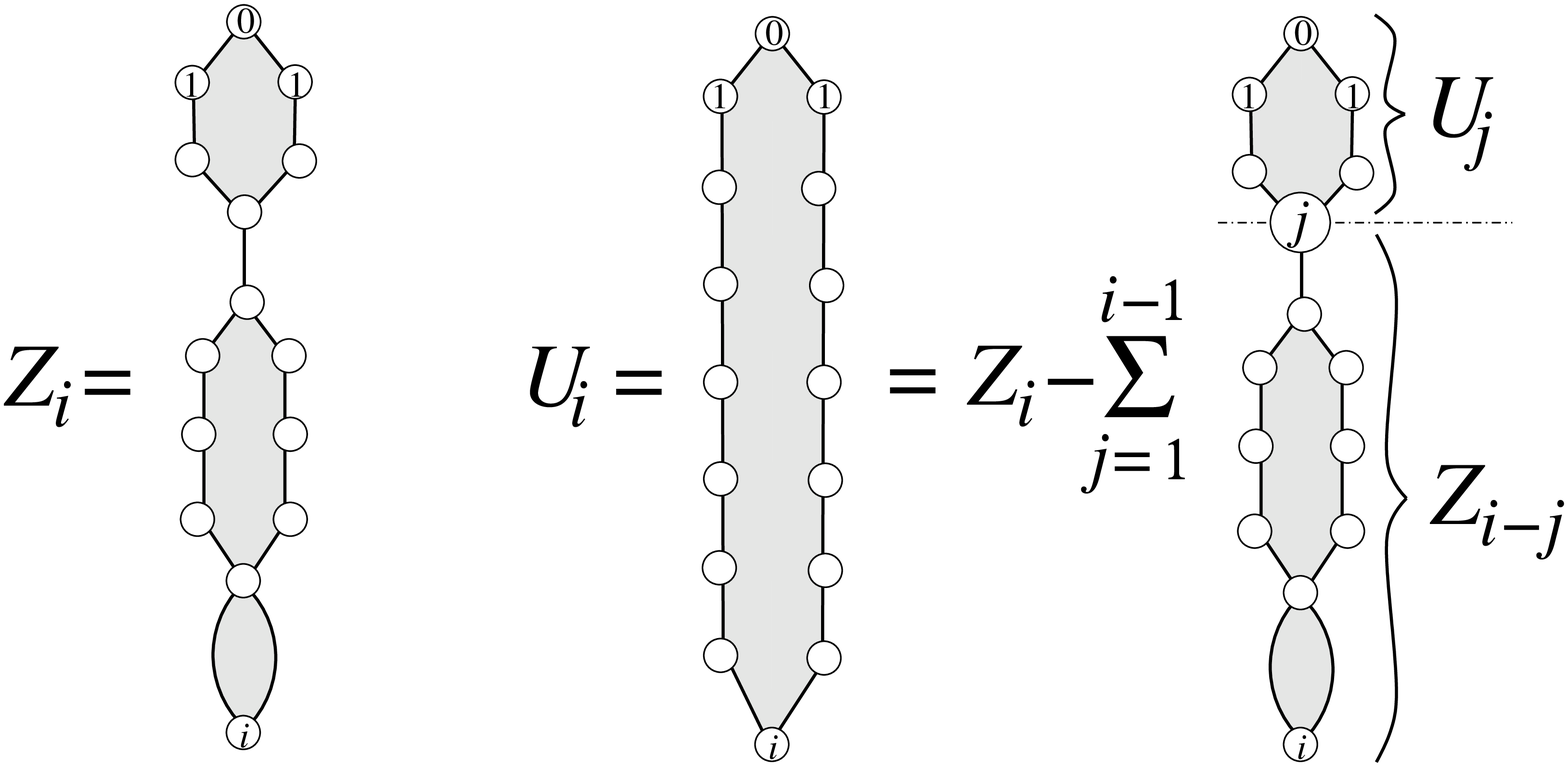}{10.cm}
\figlabel\connected
Finally, we will denote by $U_i(g)$ the generating function of 
quadrangulations with a geodesic boundary of length $2i$ made of
two geodesic paths that have no pinch point.
As we have seen, from bijection II, $U_i(g)$ is precisely the desired 
generating function
for quadrangulations with a marked geodesic of length $i$. 
From bijection III, the function $U_i(g)$ is nothing but the 
irreducible part of $Z_i(g)$
obtained inductively via the relation
\eqn\ZtoU{\eqalign{U_1(g)& =Z_1(g) = R_1(g) \cr U_i(g)& =
Z_i(g)-\sum_{j=1}^{i-1} U_j(g) Z_{i-j}(g)\ \ ,\ \ i\geq 2 \ .\cr}}
The subtraction term simply amounts to removing those quadrangulations
for which the two geodesic paths have their closest pinch point to
the origin at distance $j$, dividing the quadrangulation into a first
irreducible part ($U_j(g)$) and a remaining part with possibly
other pinch points ($Z_{i-j}(g)$), as illustrated in Fig.\connected.
\fig{The set of all quadrangulations with a geodesic boundary of length
$4$ having no pinch point, with one or two faces (in grey).}{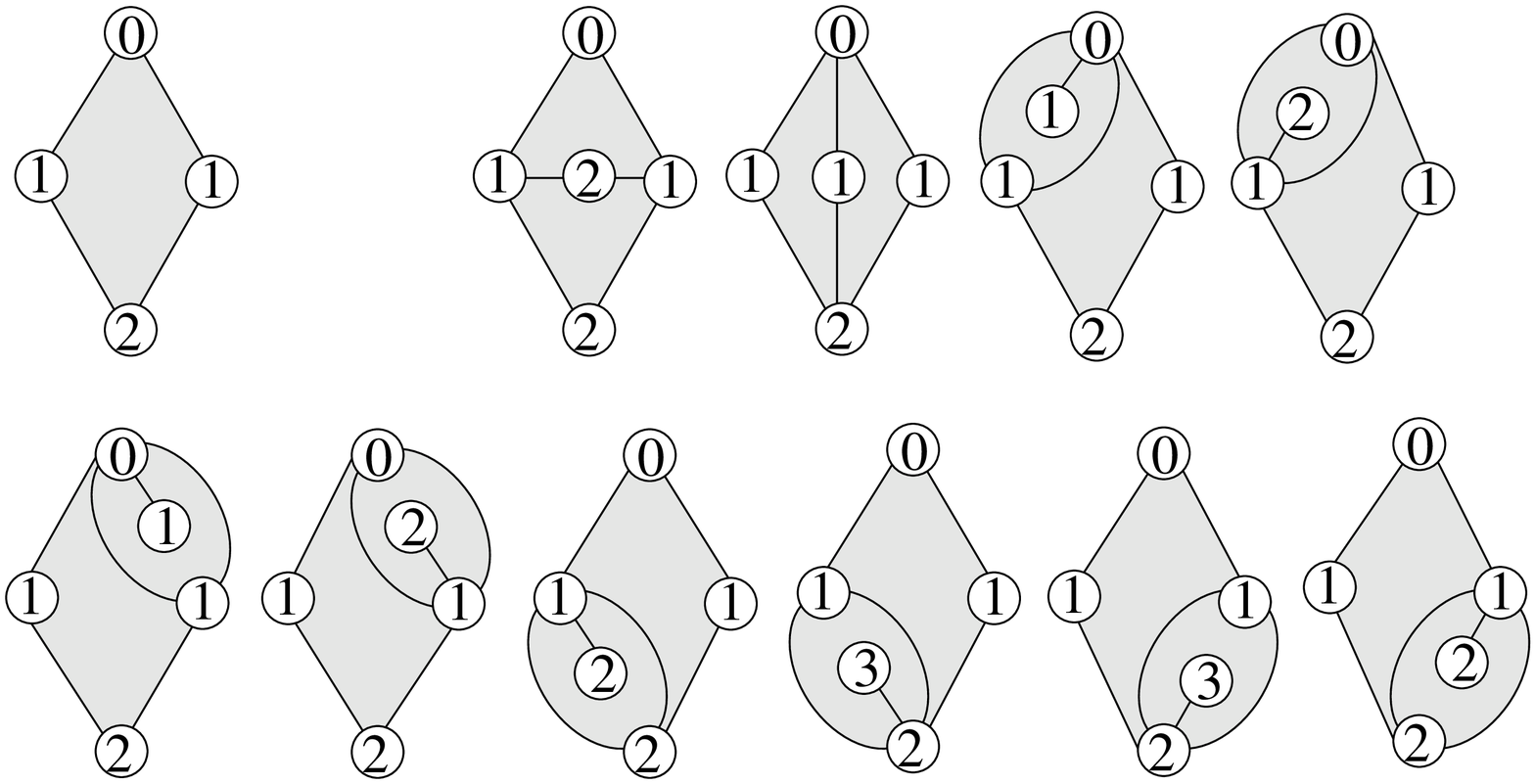}{11.cm}
\figlabel\udeux
\fig{The set of all quadrangulations of the sphere with a marked 
geodesic (thick line) of length $2$ and with one or two faces.
These configurations are in one-to-one correspondence with
those of Fig.\udeux. When drawing the quadrangulation in the plane,
we choose as external face the one lying on the left of the $0\to 1$
edge of the geodesic path.}{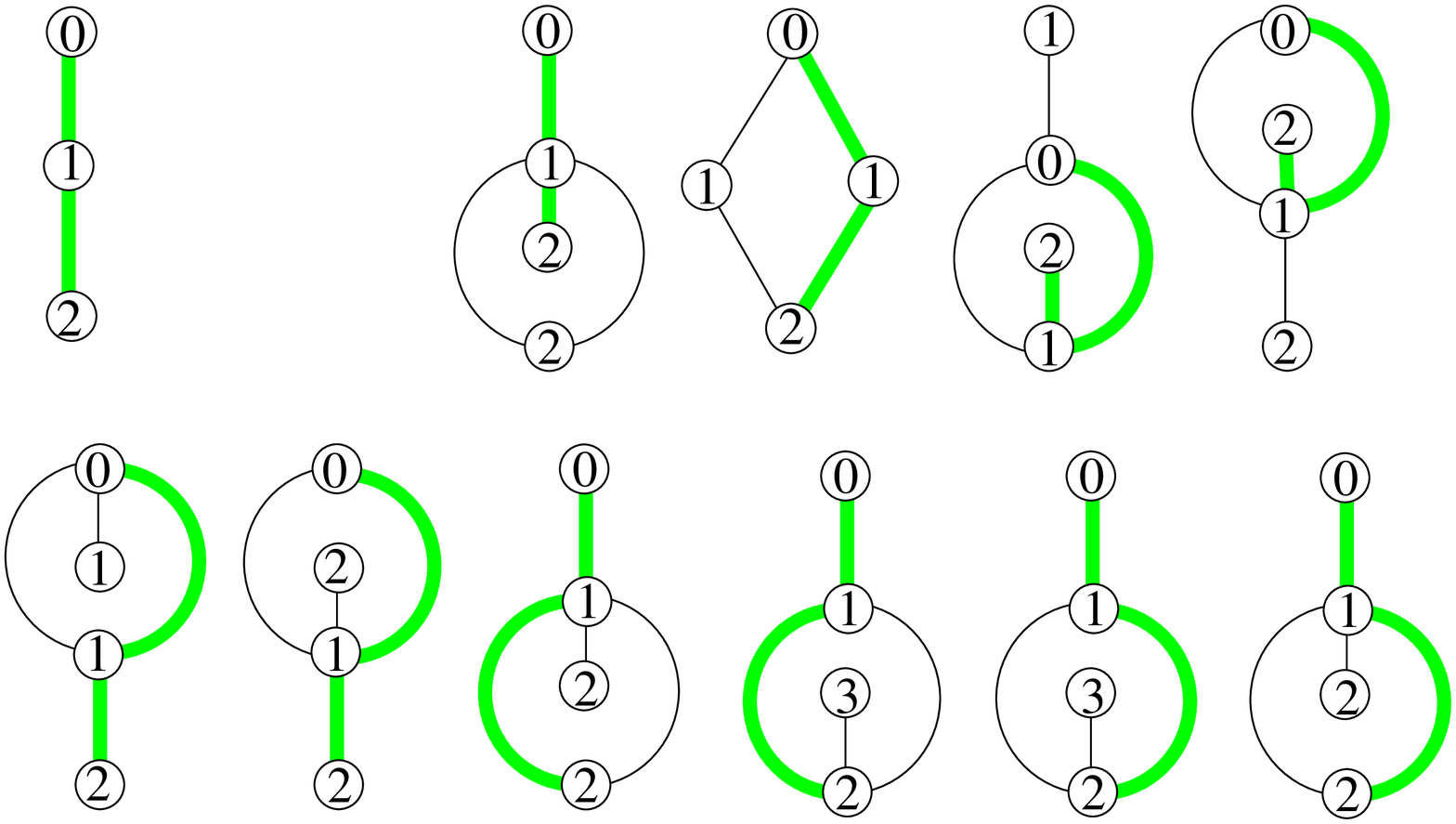}{12.cm}
\figlabel\udeuxbis
\noindent For instance, 
we have $U_2(g)=g+10 g^2 +90 g^3 +810 g^4+\ldots$, where
the first two terms correspond to the configurations of Fig.\udeux\ of
quadrangulations with a geodesic boundary of length $4$ having no pinch
point, or equivalently, to the
configurations of Fig.\udeuxbis\ of quadrangulations with a marked
geodesic of length $2$.

\subsec{Confluent geodesics} 

With the above generating functions at hand, we have also access
to generating functions for quadrangulations with several marked confluent
geodesics, namely quadrangulations with say, $k$ marked 
geodesic paths joining a common origin vertex to a common endpoint at 
distance $i$.
We shall furthermore impose that these $k$ geodesics be weakly avoiding, 
in the sense that they {\it do not cross each other}, but still can have 
common vertices or even common edges. We shall also consider the case of 
strongly avoiding 
confluent geodesics which are required to have no common vertex except for
their first and last endpoints.
In both cases, we furthermore decide to distinguish one of the geodesics
as the first one, which induces a natural linear ordering of the
geodesics.
\fig{Schematic picture of a configuration (a) with $k$ weakly avoiding 
confluent geodesics (here $k=4$). Upon unzipping (b) of the first geodesic
path (here marked by an arrow) and continuous deformation on the sphere, 
we end up with
a quadrangulation with a geodesic boundary (c) and $k-1$ geodesics in the bulk.
It can be viewed as the juxtaposition of $k$ quadrangulations with boundary
of the type depicted in Fig.\bord-(a) with the global constraint that the two
outermost boundaries cannot have common vertices.}{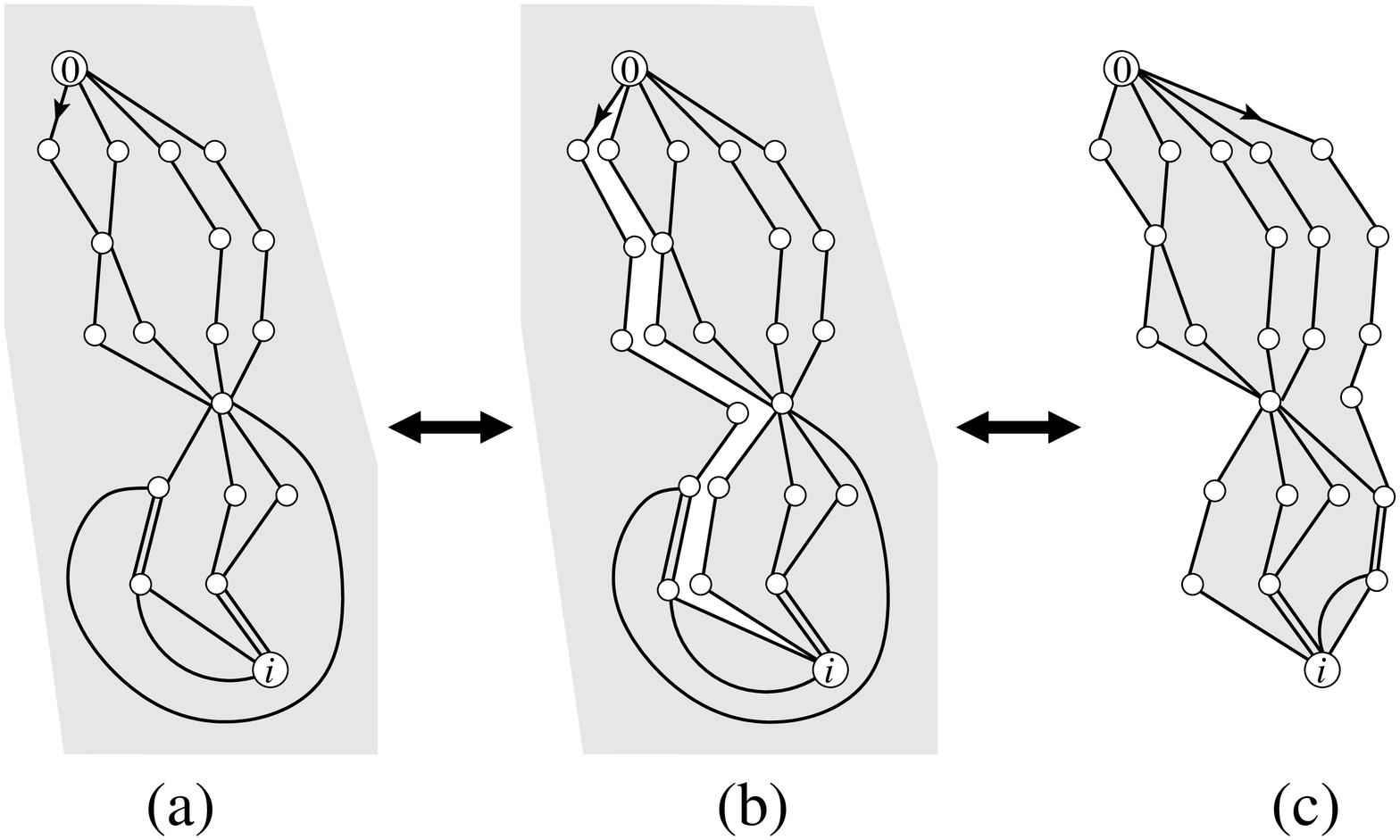}{11.cm}
\figlabel\kgeod
Now a quadrangulation with $k$ such linearly ordered weakly avoiding 
geodesics may be transformed
upon unzipping of the first geodesic and continuous deformation on the sphere
into a quadrangulations with a geodesic boundary and $k-1$ marked
geodesics in the bulk of the quadrangulation (see Fig.\kgeod). 
Note that these geodesics
may very well have common vertices or edges among themselves and with 
the boundary if the corresponding geodesics had such common vertices or edges 
on the quadrangulation. Nevertheless, there is a global requirement that the 
two boundaries themselves cannot have common vertices. Without this
requirement, the generating function would simply be 
$\big(Z_i(g)\big)^k$ by viewing the configuration as simply made of $k$ 
glued objects counted by $Z_i(g)$. To account for the global constraint, 
we again simply
have to extract the irreducible part of $\big(Z_i(g)\big)^k$, leading 
to a generating function 
$U^{(k)}_i(g)$ for quadrangulations with $k$ marked weakly avoiding 
confluent geodesics
of length $i$ given by
\eqn\Uk{\eqalign{U^{(k)}_1(g)& =\big(Z_1(g)\big)^k \cr U^{(k)}_i(g)& =
\big(Z_i(g)\big)^k-\sum_{j=1}^{i-1} U^{(k)}_j(g) \big(Z_{i-j}(g)\big)^k
\ \ ,\ \ i\geq 2 \ .\cr}}

Finally, the situation is even simpler in the case of $k$ strongly avoiding
geodesics since in this case, upon unzipping all geodesics, we obtain 
exactly $k$ quadrangulations with a geodesic boundary with no  pinch point. 
The generating 
function ${\tilde U}^{(k)}_i(g)$ for quadrangulations with $k$ marked 
strongly avoiding geodesics of length $i$ is therefore given by
\eqn\tildeUk{{\tilde U}^{(k)}_i(g)=\big(U_i(g)\big)^k\ \ .}

\newsec{Statistics of geodesics in large quadrangulations}

\subsec{From combinatorics to statistics}

Let us now see how to use the above generating functions to study
the statistics of geodesics in quadrangulations. We shall be 
interested in averaging over all quadrangulations with a 
{\it fixed number} $n$ of faces. As customary, our enumeration results
translate naturally into averages for an ensemble where
every quadrangulation is drawn with a probability proportional to its 
{\it inverse symmetry factor}, i.e.\ to the inverse of the order of its
automorphism group. In the limit $n\to \infty$ of
large quadrangulations, it is expected that the symmetry factors become
irrelevant, so that the asymptotic statistics is the same as
for the uniform probability distribution. 

\fig{The set of all pointed quadrangulations of the sphere with 
one and two faces, where we have indicated the marked vertex by a label $0$. 
For each configuration, we indicated its inverse symmetry 
factor. These add up to  ${\cal Z}_1=3/2$ for $1$ face and to ${\cal Z}_2=9/2$
for $2$ faces.}{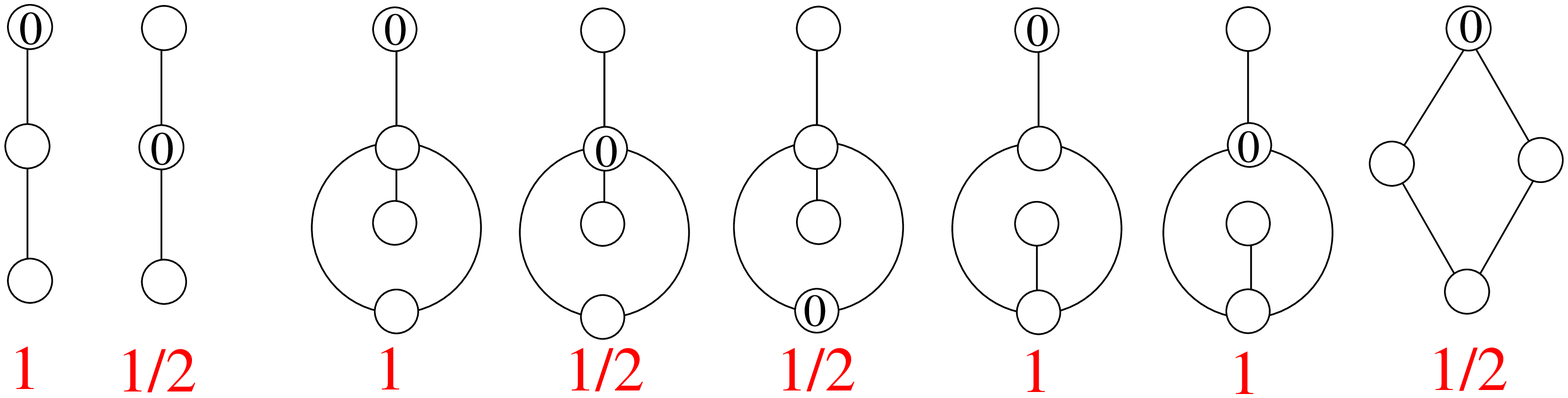}{12.cm}
\figlabel\sympointed
In our case, a natural quantity to look at is the number of geodesics 
emerging from a given origin. In this context, it is natural 
to consider the ensemble of {\it pointed quadrangulations} with $n$ faces, 
i.e.\ quadrangulations with a marked origin vertex, again weighted by
their inverse symmetry factor. The probability of a given pointed 
quadrangulation is then equal to $1/(S {\cal Z}_n)$ where $S$ is
the symmetry factor and ${\cal Z}_n$ is the partition function, which
reads  
\eqn\partit{{\cal Z}_n={3^n \over 2n} {{2n\choose n}\over n+1}} 
For instance for $n=1$ and $2$, the results ${\cal Z}_1=3/2$ and 
${\cal Z}_2=9/2$ are illustrated in Fig.\sympointed.
Note that, in contrast with rooted quadrangulations, pointed 
quadrangulations may have non-trivial symmetries, hence ${\cal Z}_n$ 
is not necessarily an integer.  

\fig{The average number of geodesics of length $2$ emerging from 
the origin of pointed quadrangulations with $1$ and $2$ faces. This
average is obtained as the sum over pointed quadrangulations of the actual 
number of geodesics (indicated in green on the right) times the symmetry 
factor of Fig.\sympointed\ (indicated in red on the left), leading
to $1$ (for $n=1$) and $10$ (for $n=2$), and dividing by the
partition function.}{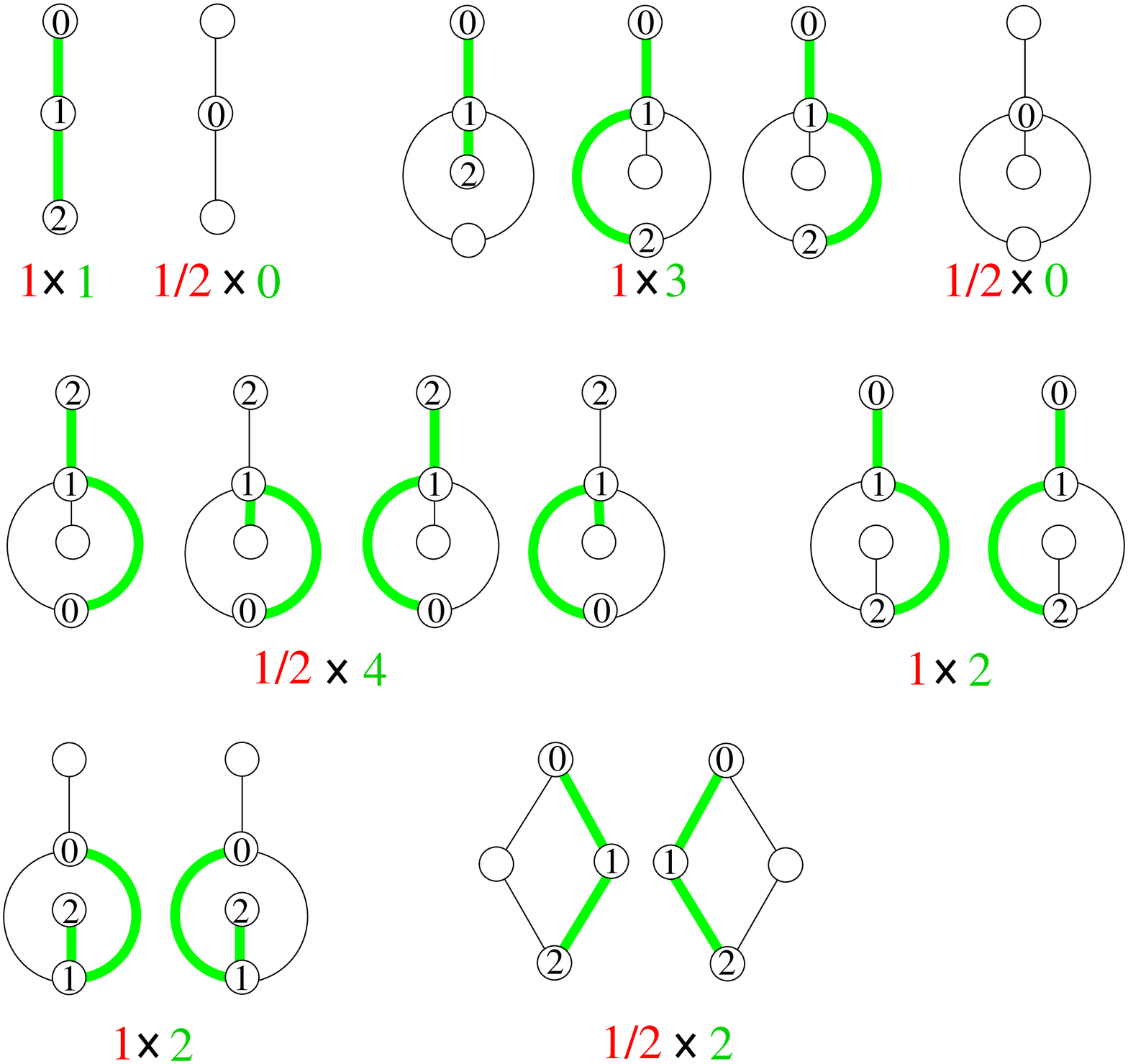}{12.cm}
\figlabel\symfact
In this ensemble, we can interpret the quantity $U_i(g)\vert_{g^n}/{\cal Z}_n$
as the average number of geodesics of length $i$ emerging from the origin.
For $i=2$, we have $U_2(g)=g+10 g^2+\ldots$, so that, for $n=1$ and $2$, 
this average number reads respectively $1/(3/2)=2/3$ and $10/(9/2)=20/9$
(see Fig.\symfact\ for an illustration).

Alternatively, we can be interested in the ensemble of quadrangulations
with two marked vertices at distance $i$ from each other, again 
weighted by symmetry factors. In that case, as mentioned above,
the partition function
is known to be $\log\big(R_i(g)/R_{i-1}(g)\big)\vert_{g^n}$. In this
ensemble, the average number of geodesics joining the two marked points
is simply $U_i(g)\vert_{g^n}$ divided by this different normalization.

\subsec{Average number of geodesics in large quadrangulations}

We are now interested in the large $n$ limit of the statistics of
geodesics. To this end, we 
first need to extract the coefficient of $g^n$ in the various generating 
functions of previous section. For an arbitrary series $X(g)$, 
this coefficient is obtained via a contour integral around $0$:
\eqn\coeffgn{X(g)\vert_{g^n}={1\over 2 {\rm i}\, \pi}\oint {dg \over g^{n+1}}
X(g)\ .}
In the case of our generating functions, the large $n$ behavior 
is governed by their singularity at the critical value $g_{\rm crit}=1/12$.
To capture this singularity, it is convenient to first change 
variable from $g$ to $V=g R(g)$, with $R(g)$ as in Eq.\resR, such that
\eqn\gtoV{g=g(V)\equiv V-3 V^2\ ,}
leading to
\eqn\coeffgnbis{X(g)\vert_{g^n}={1\over 2 {\rm i}\, \pi}\oint {dV \over \big(
V(1-3V)\big)^{n+1}} (1-6 V)\, X\big(g(V)\big)\ .}
At large $n$, the dominant behavior is given by the vicinity of saddle point
at $V_{\rm crit}=1/6$. Writing 
\eqn\Vtoxi{V={1\over 6} \left( 1+ {\rm i} {\xi \over n^{1/2}}\right)\ ,}
the contour integral over $V$ becomes at large $n$ an integral over real
values of $\xi$:
\eqn\asymp{X(g)\vert_{g^n}= {12^n \over {\rm i}\pi\, n} 
\int_{-\infty}^{+\infty} d\xi\, \xi\, e^{-\xi^2}\, 
\left(1+{\xi^4-2\xi^2\over 2 n}+{\cal O}\left(
{1 \over n^2}\right)\right)\,
X\left({1\over 12} \left(1+{\xi^2\over n}\right)\right) 
\ .}
For $X=Z_i$ or $U_i$, we will find an expansion of the form
\eqn\expX{X\left({1\over 12} \left(1+{\xi^2\over n}\right)\right) =
A + C {\xi^2 \over n} + {2\over 3}{\rm i}\, D {\xi^3 \over n^{3/2}} + 
{\cal O} \left( {1 \over n^2}\right)}
with real coefficients $A$, $C$ and $D$, and 
with a vanishing $\xi/n^{1/2}$ term. We have introduced the prefactor $2/3$
for later convenience. The even powers of $\xi$ in \expX\ do not
contribute to \asymp\ by parity and we get the leading behavior
\eqn\leadasymp{X(g)\vert_{g^n}\sim {12^n \over \pi\, n} 
\int_{-\infty}^{+\infty} d\xi\, {2\over 3}D {\xi^4\over n^{3/2}} e^{-\xi^2}
= {12^n\over 2 \sqrt{\pi} n^{5/2}} D \ .}
As explained in previous section, whenever $X(g)$ enumerates 
objects emerging from an origin, the corresponding average number 
$\langle x\rangle$ in the ensemble of pointed quadrangulations is 
obtained by dividing the above result by:
\eqn\numquad{{\cal Z}_n\sim {12^n \over 2 
\sqrt{\pi} n^{5/2}}\ ,}
so that the asymptotic average simply reads
\eqn\aveform{\langle x\rangle = D\ .}

Using the explicit form \resZ\ for $Z_i$, we find immediately the expansion
\eqn\devZi{\eqalign{Z_i&= 
A_i + C_i {\xi^2 \over n} + 
{2\over 3}{\rm i}\, D_i {\xi^3 \over n^{3/2}} + \ldots \cr
{\rm with}\ \ & A_i= {2^i (i+3)\over 3(i+1)} \cr
& C_i= {2^i i(i+3)(i^2+6i+3)\over 30(i+1)}\cr & D_i=
{2^i i (i+2)(i+3)(i+4)(3 i^2 +12 i+13)\over 420 (i+1)}\cr}}
from which we obtain the large $n$ asymptotic behavior 
\eqn\asympZi{Z_i(g)\vert_{g^n}= {12^n\over \sqrt{\pi} n^{5/2}} 
{2^i i (i+2)(i+3)(i+4)(3 i^2 +12 i+13)\over 840 (i+1)} }
As for $U_i$, we have a similar expansion 
\eqn\devUi{U_i= \alpha_i + \gamma_i {\xi^2 \over n} +
 {2\over 3}{\rm i}\, \delta_i 
{\xi^3 \over n^{3/2}} + \ldots}
where the coefficient $\delta_i$ gives the large $n$ behavior 
of $U_i$ via the general relation \leadasymp. In particular, from 
\aveform, 
{\it the average number of geodesic paths of length $i$ emerging
from the origin vertex in the ensemble of pointed quadrangulations of
fixed large size is directly given by $\delta_i$}.
The coefficients $\alpha_i$, $\gamma_i$ and $\delta_i$ can be
related to $A_i$, $C_i$ and $D_i$ via the relation \ZtoU, namely
\eqn\Atoalpha{\eqalign{\alpha_i & =A_i-\sum_{j=1}^{i-1}\alpha_j A_{i-j} \cr
\gamma_i& =C_i-\sum_{j=1}^{i-1} \left(\alpha_j C_{i-j}+\gamma_j A_{i-j}
\right) \cr
\delta_i& =D_i-\sum_{j=1}^{i-1} \left(\alpha_j D_{i-j}+\delta_j A_{i-j} \right)
 .\cr}}
These relations are better expressed by introducing generating
functions for these coefficients:
\eqn\gencoeff{\eqalign{
\hat A(t) & \equiv \sum_{i=1}^\infty A_i t^i,\qquad 
\hat C(t)\equiv \sum_{i=1}^\infty C_i t^i, \qquad 
\hat D(t)\equiv \sum_{i=1}^\infty D_i t^i \cr 
\hat \alpha(t)& \equiv \sum_{i=1}^\infty \alpha_i t^i, \qquad 
\hat \gamma(t) \equiv \sum_{i=1}^\infty \gamma_i t^i, \qquad 
\hat \delta(t) \equiv \sum_{i=1}^\infty \delta_i t^i \cr}} 
as they translate into:
\eqn\Dtodelta{\hat \alpha(t)  = {\hat A(t) \over 1+\hat A(t)}, \quad
\hat \gamma(t) = { \hat C(t) \over (1+\hat A(t))^2} , \qquad
\hat \delta(t) = { \hat D(t) \over (1+ \hat A(t))^2}}
{}From the explicit values of $A_i$, $C_i$ and $D_i$, we can get explicit values
for all these generating functions, leading to:
\eqn\expli{\eqalign{\hat \alpha(t) &= {t(6t-2)-(1-2t)\log(1-2t) \over
t-(1-2t) \log(1-2t)} \cr 
\hat \gamma(t)&= {3t \left(2t \left( 20 t^3 -34 t^2 +17 t-1\right) -(1-2t)^4
\log(1-2t)\right) \over 5(1-2t)^2 \left(t-(1-2t)\log(1-2t)\right)^2}
\cr
\hat \delta(t)&= {3t \left(2t \left(3+177 t 
-412 t^2 +708 t^3 -624 t^4 +
224 t^5\right) +3\left(1-2t\right)^6 \log(1-2t)\right)
\over 70 \left(1-2t\right)^4 \left(t-(1-2t) \log(1-2t) \right)^2}\cr
&= 4 t +{80 \over 3}t^2+132 t^3+{366208
    \over 675}t^4+{3998176 \over 2025}t^5+{\cal O}\left(t^6\right)\ .\cr }}
The function $\hat \delta(t)$ is the generating function for the average number 
$\delta_i$ of geodesics of length $i$ emerging 
from the origin vertex in a large pointed
quadrangulation. From the last line of \expli, we directly read of the 
values of this average number for a length $i=1,2,\cdots,5$.

We can easily estimate the large $i$ behavior of $\alpha_i$, $\gamma_i$
and $\delta_i$
by looking at the singularity of $\hat \alpha(t)$, $\hat \gamma(t)$ 
and $\hat \delta(t)$ 
for $t\to 1/2$. Writing
$t=1/2(1- \eta)$, we have the small $\eta$ behavior
\eqn\smalleta{\eqalign{
\hat \alpha\left(
{1\over 2}(1-\eta)\right)& =1-3\eta-6 \eta^2 \log(\eta)+\ldots 
\cr \hat \gamma\left({1\over 2}(1-\eta)\right)&\sim {9\over 5 \eta^2}\cr
\hat \delta\left({1\over 2}(1-\eta)\right)&\sim {54\over 7 \eta^4}\cr} }
Note that this last behavior follows immediately from the small $\eta$ behaviors
$\hat A(t)\sim 1/(3\eta)$ and $\hat D(t)\sim 6/(7 \eta^6)$ and 
from the relation
\Dtodelta. From \smalleta, we deduce:
\eqn\asydelt{
\alpha_i \buildrel {i\to \infty} \over \sim
2^i \, {12 \over i^3}, \qquad
\gamma_i \buildrel {i\to \infty} \over \sim
2^i \, {9 \over 5}\ i, \qquad
\delta_i \buildrel {i\to \infty} \over \sim
2^i \, {9 \over 7}\ i^3 \ .}
From the discussion of previous section, to go to the ensemble of 
quadrangulations with two marked vertices at distance
$i$, we simply have to gauge $\delta_i$ by the average number of points
at distance $i$ from the origin in large pointed quadrangulations. This
number is given by $(3/7) i^3$ for large $i$ (see for instance Eq.(5.2) in 
Ref.\MOB), hence we deduce that the 
average number $\langle {\cal G}\rangle_i$ of 
geodesics joining two fixed points 
at distance $i$ from each other is simply given by
\eqn\avgeo{\langle {\cal G}\rangle_i\sim 3\times 2^i} 
at large large $i$. The leading exponential $i$-dependence can 
be heuristically understood as follows: note that choosing
a particular geodesic leading to a given vertex simply amounts,
when going backwards from that vertex, to choose recursively, for
each vertex at distance $i$ a predecessor vertex at distance $i-1$,
until we eventually reach distance $0$. Now one can easily show that, for
a vertex at distance $i\gg 1$ from the origin of a random quadrangulation,
its average number of neighbors at distance $i-1$ is $2$.  
There are thus on average $2$ choices for each of the $i$ steps leading
back to the origin, which explains the $2^i$ factor by neglecting 
correlations. The prefactor $3$ precisely accounts for these correlations
as well as for the finite $i$ corrections to the asymptotic number
of choices $2$.

\subsec{Average number of confluent geodesics} 
We can as well obtain the average number of $k$-tuples of 
(linearly ordered) weakly avoiding confluent geodesics of length $i$ by 
considering the expansion:
\eqn\devZik{\eqalign{\left(Z_i\right)^k &= 
A^{(k)}_i + C^{(k)}_i {\xi^2 \over n} + 
{2\over 3}{\rm i}\, D^{(k)}_i {\xi^3 \over n^{3/2}} + \ldots \cr
{\rm with}\ \ & A^{(k)}_i= \left(A_i\right)^k \cr
& C^{(k)}_i= k\ \left(A_i\right)^{k-1}\, C_i\cr
& D^{(k)}_i= k\ \left(A_i\right)^{k-1}\, D_i\ ,\cr}}
where the coefficients $A_i$, $C_i$ and $D_i$ are given by Eq.\devZi.
In particular, the generating functions $\hat A^{(k)}(t)\equiv 
\sum_{i\geq 1} A^{(k)}_i t^i$,
$\hat C^{(k)}(t)\equiv \sum_{i\geq 1} C^{(k)}_i t^i$ 
and $\hat D^{(k)}(t)\equiv \sum_{i\geq 1} D^{(k)}_i t^i$ 
now have a singularity at $t=1/2^k$. 
 
Similarly, the function $U_i^{(k)}$ has the expansion 
\eqn\devUik{U_i^{(k)}= 
\alpha^{(k)}_i + \gamma^{(k)}_i {\xi^2 \over n} + 
{2\over 3}{\rm i}\, \delta^{(k)}_i {\xi^3 \over n^{3/2}} + \ldots }
where, as before, the coefficients $\alpha^{(k)}_i$, $\gamma^{(k)}_i$
and $\delta^{(k)}_i$ are determined through their generating functions:
\eqn\genek{\eqalign{\hat \alpha^{(k)}(t)& \equiv 
\sum_{i\geq 1} \alpha^{(k)}_i t^i= {\hat A^{(k)}(t)\over 1+\hat A^{(k)}(t)}\cr
\hat \gamma^{(k)}(t)&\equiv \sum_{i\geq 1} \gamma^{(k)}_i t^i =
{\hat C^{(k)}(t)\over \left( 1+\hat A^{(k)}(t)\right)^2} \cr 
\hat \delta^{(k)}(t)&\equiv \sum_{i\geq 1} \delta^{(k)}_i t^i =
{\hat D^{(k)}(t)\over \left( 1+\hat A^{(k)}(t)\right)^2} \ .\cr }}
Writing
$t=1/2^k\, (1-\eta)$ and expanding at small $\eta$, we have now the
dominant behaviors $\hat A^{(k)}(t)\sim 1/(3^k \eta)$ and 
$\hat D^{(k)}(t)\sim 6k/(3^{k-1}\,7\eta^6)$, leading directly to 
$\hat \delta^{(k)}(t) \sim (k\,3^{k-1}\, 54)/(7\eta^4)$. From 
this singularity, we immediately deduce the large $i$ behavior of 
$\delta_i^{(k)}$:
\eqn\asydeltk{\delta_i^{(k)} \buildrel {i\to \infty} \over \sim
\left(2^i\right)^k \, k\, 3^{k-1}\, {9 \over 7}\ i^3 }
which gives the average number of $k$-tuples of (linearly ordered) 
weakly-avoiding confluent
geodesics of length $i$ emerging from the origin of a large pointed 
quadrangulation. Again, when compared with the average number 
$(3/7) i^3$ of vertices at distance $i$ from this origin, this gives 
an average number
\eqn\avekgeo{\langle {\cal G}_k \rangle_i \sim k (3 \times 2^i)^k}
of such $k$-tuples between the two marked points in the ensemble of 
quadrangulations with a large fixed size and two marked points at 
distance $i$ from each other.

Let us make a few remarks on this particularly simple result. 
First, the leading exponential $i$-dependence $2^{ik}$ is surprisingly
the same as for $k$ independent geodesics. One could have expected
a reduction of the entropy factor due to the constraint of
non-intersection between geodesics. Indeed, as we shall see later, weakly
avoiding confluent geodesics have a number of contacts which is proportional
to their length. This apparent paradox can be explained 
as follows: as before, we can estimate the number of geodesics by looking
at them backwards and choosing recursively for each vertex at distance $i$
one of its predecessors at distance $i-1$. For a vertex visited by a single 
geodesic, this gives $2$ choices on average as before. When a number $\ell$ 
of geodesics come in contact at a vertex with $p$ predecessors, 
the number of choices is now ${p+\ell-1 \choose  \ell}$ (instead of $p^\ell$ 
for freely intersecting geodesics). Now the probability of having $p$ 
predecessors for a vertex at a large distance from the origin can be easily 
shown to be $1/2^p$ (with $p$ varying from $1$ to $\infty$), leading
to an average value $\langle {p+\ell-1 \choose  \ell}\rangle = 2^\ell$, 
i.e.\ the same contribution as if the $\ell$ geodesics were visiting
independent vertices. In other words, the reduction of choices among 
predecessors due to non-intersection is exactly compensated by the effect 
of correlations for the number of predecessors at a given vertex.

Even more surprising, the sub-leading prefactor $3$ in Eq.\avgeo\ is simply
raised to its $k$-th power as if the effect of small $i$ corrections
and correlations along the geodesics could be treated independently
for each geodesic path. Finally, we have in Eq.\avekgeo\ a global
prefactor $k$ which can be understood as a spontaneous symmetry breaking
effect as follows: the $k$ confluent geodesics are linearly ordered 
and any pair of neighboring geodesics 
encloses a particular domain of the quadrangulation, with $k$ such
domains. As we shall see,
for quadrangulations with a large number $n$ of faces, only one of these 
$k$ domains contains most of the area of the quadrangulation,
i.e.\ has a number of faces close to $n$, while all the other domains have 
an area negligible with respect to $n$ at large $n$. There are precisely $k$ 
choices for the domain in which most the area will lie, hence the 
global prefactor $k$.

\subsec{Other statistical properties of confluent geodesics}

By slight refinements of the above arguments, we have access to more
involved statistical properties of confluent geodesics. For instance,
we may compute the asymptotic (large $n$) average number $\langle c \rangle_i$ 
of contacts between two weakly avoiding confluent geodesics of length $i$,
with result:
\eqn\contact{\langle c \rangle_i \sim {i\over 3}}
at large $i$. A proof of this result is presented in Appendix A using
the continuous formalism of Section 4. 
As announced, this number is proportional to the length of the geodesics
which indicates that the two geodesics ``stick" to one another.
\fig{Schematic picture of a configuration (a) with two weakly avoiding
confluent geodesics. The two geodesics delimit two (possibly disconnected)
complementary domains (here in dark and light) lying between them.
The same configuration (c) obtained upon unzipping of one geodesic path 
(b) and continuous deformation on the sphere.}{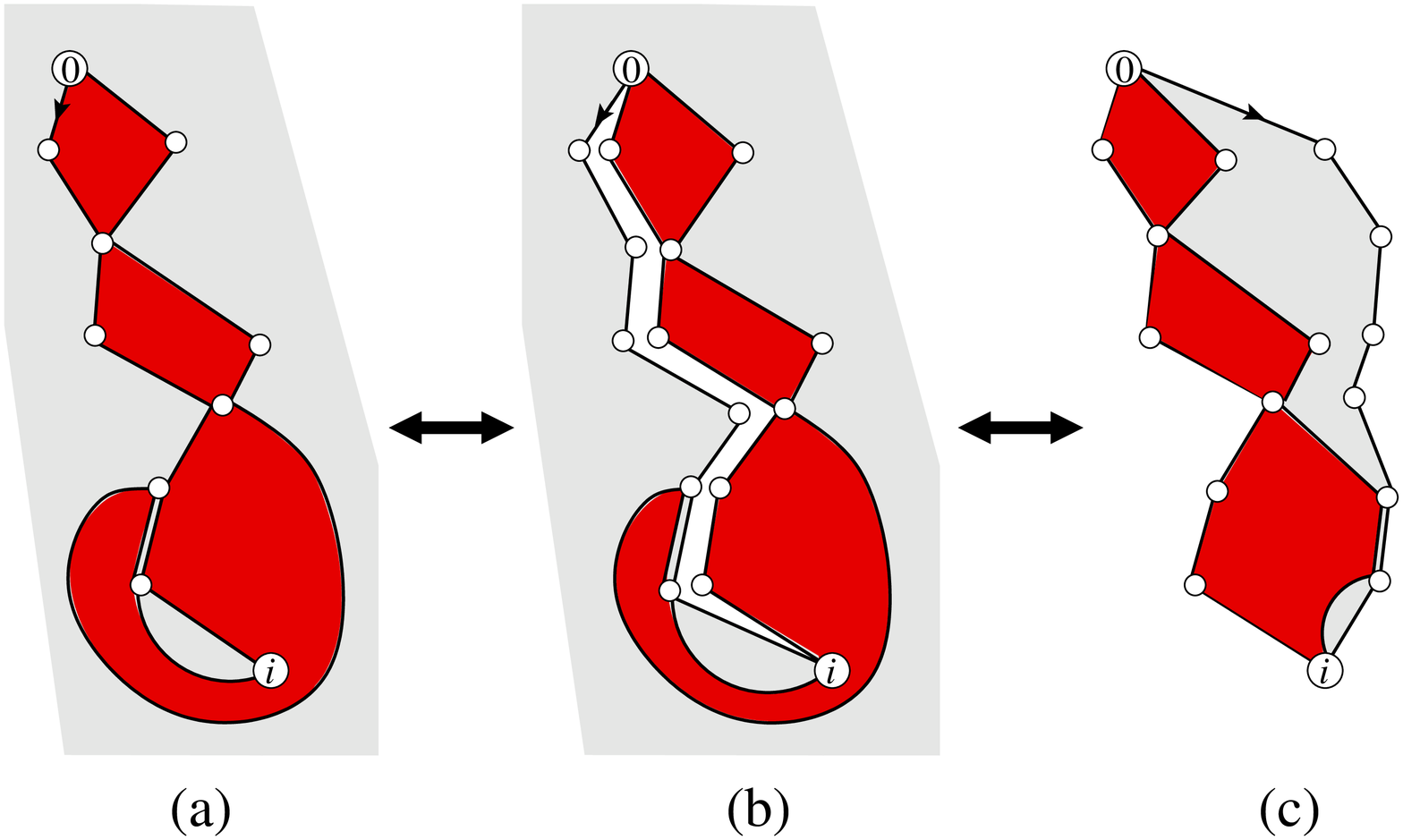}{10.cm}
\figlabel\masses
Another quantity of interest is the area (number of faces) enclosed by 
these two geodesics.  More precisely, let $n_1$ and $n_2$ be the areas
of the two domains delimited by the geodesics in the quadrangulation,
with $n_1+n_2=n$ (see Fig.\masses\ for an illustration).
By symmetry, we have $\langle n_1\rangle =\langle n_2\rangle =n/2$ and
a good measure of the dispatching of the area is given by the correlation
$\langle n_1 n_2 \rangle$. For two geodesics of length $i$, we find:
\eqn\dispat{\langle n_1 n_2 \rangle_i \sim n \theta_i}
for $n \to \infty$ and $i$ finite, with a finite coefficient $\theta_i$. 
This implies a very dissymmetric dispatching where the smaller domain has 
an average area of order $\theta_i$ only while almost all of the area lies 
in the larger domain. For large $i$, we obtained the leading behavior:
\eqn\lartheta{\theta_i \sim {27\over 100} i^3\ .}
A derivation of this result is given in Appendix B by use of the
continuous framework of Section 4.
The above dissymmetry generalizes easily to the case of domains delimited
by $k$ confluent geodesics. We find that only one of these domains has 
an extensive area, meaning that the geodesics remain close to each other. 

In view of the above results, we expect very different statistics
for strongly avoiding confluent geodesics where contacts
are forbidden. The asymptotic number of $k$-tuples of such
geodesics of length $i$ is obtained from the expansion of 
${\tilde U}^{(k)}_i$, namely:
\eqn\devUikbis{\eqalign{{\tilde U}^{(k)}_i &= 
{\tilde \alpha}^{(k)}_i + {\tilde \gamma}^{(k)}_i {\xi^2 \over n} + 
{2\over 3}{\rm i}\, {\tilde \delta}^{(k)}_i {\xi^3 \over n^{3/2}} + \ldots \cr
{\rm with}\ \ 
& {\tilde \alpha}^{(k)}_i= \left(\alpha_i\right)^k \cr
& {\tilde \gamma}^{(k)}_i=  k\,\left(\alpha_i\right)^{k-1}\, \gamma_i\cr 
& {\tilde \delta}^{(k)}_i=  k\,\left(\alpha_i\right)^{k-1}\, \delta_i \ .\cr}}
{}From the large $i$ behavior \asydelt, we get
\eqn\asytilddelt{{\tilde \delta}^{(k)}_i \buildrel {i\to \infty} \over \sim
k \left(3\times 2^i\right)^k  {3 \times 4^{k-1} \over 7}\ i^{6-3k}\ .}
consisting of the same degeneracy factor $k(3\times 2^i)^k$ as in Eq.\avekgeo\
multiplied by a factor $(3\, 4^{k-1} / 7) \ i^{6-3k}$ which we interpret
as counting the average number of vertices that are indeed reachable by $k$ 
strongly avoiding geodesics. For $k>2$, this factor goes to zero at large
$i$, which indicates that distant vertices cannot be 
reached by more than $2$ strongly avoiding geodesics. For $k=2$, we have 
for each (large) value $i$ of the distance, an average number $12/7$ of 
vertices reachable from the origin by $2$ strongly avoiding geodesics. 
We shall call such vertices {\it exceptional points}. 

Let us now consider the areas $n_1$ and $n_2$ of the two domains
delimited by $2$ strongly avoiding confluent geodesics of
length $i$. Again we want to compute the correlation 
$\langle n_1 n_2\rangle$. This is done by considering
the generating function $\left(g {d\over dg} U_i(g)\right)^2$ as the 
action of $g {d\over dg}$ amounts precisely to giving a weight proportional
to the area. When going to the $\xi$ variable, we have at leading
order in $n$ that $g {d\over dg}={n\over 2 \xi} {d\over d \xi}$ hence,
from the expansion \devUi, we get the expansion
\eqn\expdg{\left(g {d\over dg} U_i(g)\right)^2
=\gamma_i^2 
+ 2 {\rm i} \, \gamma_i\delta_i {\xi \over n^{1/2}}+\ldots}
Performing the integral over $\xi$, the first term vanishes
by parity and we get the dominant behavior
\eqn\leaddg{\left.\left(g {d\over dg} U_i(g)\right)^2 
\right\vert_{g^n}
\sim {12^n\over
\sqrt{\pi} n^{3/2}} \gamma_i\delta_i\ .}
This is to be compared with the asymptotic behavior
\eqn\leadUg{\left.\left(U_i(g)\right)^2 \right\vert_{g^n}\sim {12^n\over
2 \sqrt{\pi} n^{5/2}} \delta^{(2)}_i = {12^n \over
\sqrt{\pi} n^{5/2}} \alpha_i \delta_i\ .}
Taking the ratio of \leaddg\ and \leadUg, we get the asymptotic
average
\eqn\masmas{\langle n_1 n_2 \rangle_i = n\, {\gamma_i\over \alpha_i} 
\buildrel {i\to \infty} \over \sim {3\over 20 } n\, i^4\ .}
Again, almost all of the area is carried by a single domain, while
the smaller domain now has an area of order $i^4$. 
This result is valid in the limit $n\to \infty$ and $i$ finite but large.
It will be extended in the next section to the scaling limit of large 
quadrangulations, which is obtained by taking $i\propto n^{1/4}$. 
In this limit, we have $i^4 \propto n$, which means that the smaller 
area is of the same order as the larger one, i.e.\ the two domains
contain a finite fraction of the total area. 

\newsec{Continuum limit}

\subsec{Reminders}
So far the results which we presented were obtained in the limit of
a size $n$ of the quadrangulations tending to infinity, keeping a finite 
value for the length $i$ of the geodesics. On the other hand, it is well
known that an interesting scaling limit may be obtained by letting both $n$ 
and $i$ tend to infinity, while keeping the ratio $i/n^{1/4}$ fixed.
\fig{The distribution function $\Phi(r)$ measuring the probability that
a vertex be at a rescaled distance less than $r$ in the scaling limit
of large quadrangulation. Its derivative $\rho(r)$ is the density of
vertices at rescaled distance $r$.}{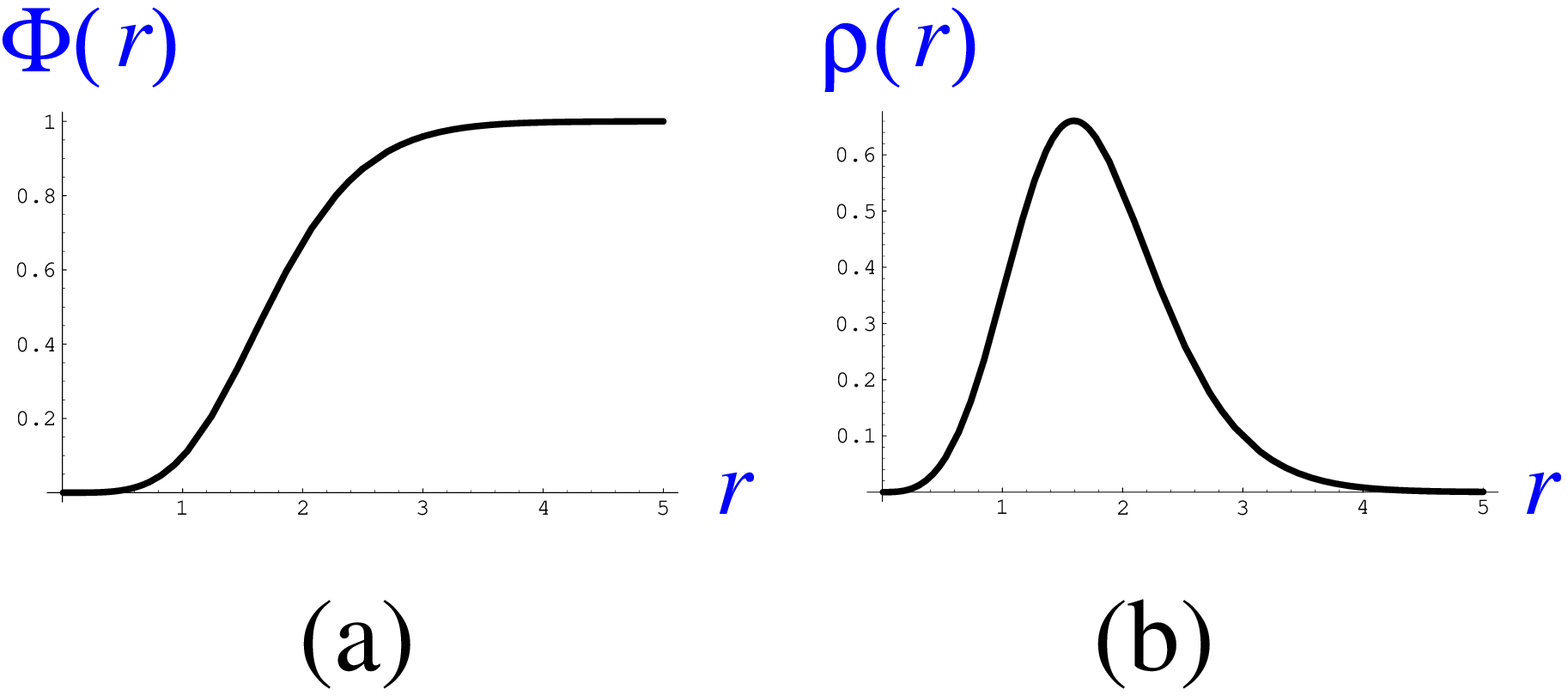}{13.cm}
\figlabel\rhophi
As a reminder, let us first consider the average number of vertices within
a ball of radius $i=r\, n^{1/4}$ centered at a given origin in a random 
quadrangulation of size $n$. In the limit $n\to \infty$, this average
number behaves as $n \Phi(r)$, with $\Phi(r)$ a finite scaling function.
This function was computed in Ref.\GEOD, with the result:
\eqn\expliphi{\Phi(r)= {4\over \sqrt{\pi}} \int_0^\infty d\xi\, \xi^2\,  
e^{-\xi^2} \left(1-6 {1-\cosh(r \sqrt{3 \xi}) \cos(r \sqrt{3 \xi})\over
\left(\cosh(r \sqrt{3 \xi})-\cos(r \sqrt{3 \xi})\right)^2}\right)\ .}
It increases from $\Phi(0)=0$ to $\Phi(+\infty)=1$ as it should and its 
derivative
\eqn\explirho{\rho(r)\equiv \Phi'(r)}
may be interpreted as the average density of vertices at a
rescaled distance $r$ from a given origin. Both functions $\Phi(r)$
and $\rho(r)$ are plotted in Fig.\rhophi\ for illustration. 
We have in particular the asymptotic behaviors:
\eqn\asyrho{\eqalign{
\rho(r) & \buildrel {r\to 0} \over \sim {3\over 7} r^3 \cr
\rho(r) & \buildrel {r\to \infty} \over \sim e^{-3 (3/8)^{2/3} r^{4/3}}
\ .\cr}}
The rest of this section is devoted to the computation of similar 
scaling functions characterizing the statistics of geodesics in
the same scaling limit.

\subsec{Distribution of geodesics of length $r$}

Let us first consider the case of quadrangulations of size $n$, with
a single marked geodesic of length $i$, as counted by $U_i(g)\vert_{g^n}$. 
To estimate this quantity in the scaling limit, we first evaluate 
$Z_i(g)\vert_{g^n}$ from its explicit form \resZ. For $i=r\, n^{1/4}$, we 
have the expansion
\eqn\expanlimit{\eqalign{ & {1\over 2^i} Z_i\left({1\over 12}
\left(1+{\xi^2\over n}\right)\right) =  {1\over 3} +{{\cal F}(r,\xi) \over
n^{1/4}} +{\cal O}\left({1\over n^{1/2}}\right) \cr
& {\rm with}\ \ {\cal F}(r,\xi)= \sqrt{-{2 \over 3}\, {\rm i}\, \xi}\ 
\coth\left(r\ \sqrt{-{3 \over 2}\, {\rm i}\, \xi}\right)
+{{\rm i}\, r \, \xi\over 3} \ ,\cr}}
which, upon substituting in \asymp, gives the asymptotic 
behavior
\eqn\scalingZ{\left.{Z_i\over 2^i}\right\vert_{g^n}
\buildrel {n\to \infty} \over \sim {12^n \over \pi n^{5/4}}
\int_0^{\infty} d\xi\, \xi\, e^{-\xi^2}
\left( {2 r\xi \over 3} - 2 \sqrt{\xi \over 3} {\sinh(r \sqrt{3 \xi})-
\sin(r \sqrt{3 \xi})\over \cosh(r \sqrt{3 \xi})-\cos(r \sqrt{3 \xi})}\right)\ .}
In particular, for $r\to 0$, the integral behaves as $\sqrt{\pi } r^5/ 280$, 
leading to $(Z_i/2^i) \sim 12^n/(\sqrt{\pi} n^{5/4}) \times (r^5/ 280)$, 
consistent with Eq.\asympZi\ in the regime $1\ll i \ll n^{1/4}$.

As before, we may obtain $U_i$ from $Z_i$ via equation \ZtoU, which we
rewrite as
\eqn\reZtoU{{\hat U}(t)={{\hat Z}(t) \over 1+{\hat Z}(t)} }
upon introducing the generating functions
\eqn\genUZ{{\hat Z}(t)\equiv \sum_{i=1}^\infty Z_i t^i, \qquad
{\hat U}(t)\equiv \sum_{i=1}^\infty U_i t^i,}
with a weight $t$ per geodesic step.
The scaling limit $i\propto n^{1/4}\to \infty$ is now captured by 
taking 
\eqn\scalt{t={1\over 2}\, e^{-s n^{-1/4}}\ .}
In this regime, we cannot simply plug the asymptotic form \expanlimit\ of
$Z_i$ in $\hat Z(t)$ as ${\cal F}(r,\xi)$ has a pole at $r=0$. A simple
way to work around this problem is to subtract from $Z_i$ its critical 
value $Z_i(1/12)=A_i$, as defined in Section 3.2, and to write
\eqn\elevexp{(Z_i-A_i)t^i={e^{-s\, r} \over n^{1/4}}\left({\cal F}(r,\xi)
-{2\over 3\, r}\right)+{\cal O}\left({1\over n^{1/2}}\right)\ .}
Noting that the right hand side has no more pole, the
sum over $i$ can be approximated by an integral over $r$, leading to:
\eqn\laplaceZ{{\hat Z}(t)-\hat A(t)
= \int_0^{\infty} dr\, e^{-s\, r}\left({\cal F}(r,\xi)-{2\over 3\, r}\right)
+{\cal O}\left({1\over n^{1/4}}\right)\ ,}
with, as before, $\hat A(t)=\sum A_i t^i$. From the explicit form \devZi,
we find the explicit expansion of $\hat A(t)$:
\eqn\exat{\hat A(t)={n^{1/4}\over 3s} -{2\over 3} \log \left(
{s\over n^{1/4}}\right) -{5\over 6}+{\cal O}\left({1\over n^{1/4}}\right)\ . }
Plugging these expansions into \reZtoU, we deduce the expansion of $\hat U(t)$
\eqn\laplaceU{\eqalign{{\hat U}(t)&
= 1-{3 s\over n^{1/4}} +{s^2 \over n^{1/2}}\left( -6 \log 
\left({s\over n^{1/4}} \right) +{3\over 2} + 9
\int_0^{\infty} dr\, e^{-s\, r}  \left(
{\cal F}(r,\xi) -{2\over 3 \, r} \right)\right) +\ldots\cr
& = \hat \alpha(t) +{1\over n^{1/2}} \int_0^\infty dr\, e^{-s\, r}
\left(9 {\cal F}''(r, \xi)-{12\over r^3}\right)+\ldots \cr}
}
where we have identified the expansion of $\hat \alpha(t)$, as defined 
in Section 3.2, and with 
${\cal F}''\equiv \partial^2{\cal F}/\partial r^2$.
Going back to $U_i(g)$, this gives an expansion
\eqn\expanlimU{{1\over 2^i} (U_i-\alpha_i) = {1\over n^{3/4}}
\left(9 {\cal F}''(r,\xi) -{12 \over r^3}\right)
+{\cal O}\left({1\over n}\right)\ ,}
which, from the asymptotics $\alpha_i \sim 12/i^3$ of Eq. \asydelt, yields
the scaling form
\eqn\scforU{{1\over 2^i} U_i\left({1\over 12}\left(1+{\xi^2 \over n}\right)
\right)= {9 {\cal F}''(r,\xi) \over n^{3/4}}+
{\cal O}\left({1\over n}\right) \ . }
Surprisingly, when comparing with Eq.\expanlimit, we see that, in the scaling 
limit, extracting the irreducible part simply amounts to differentiating 
twice with respect to $r$.
Upon substitution of \scforU\ in Eq.\asymp, we obtain the desired 
asymptotic scaling behavior
\eqn\scalU{\eqalign{\left.{U_i\over 2^i}\right\vert_{g^n} &
\buildrel {n\to \infty} \over \sim {12^n \over \pi n^{7/4}}\ 9 {\partial^2 \over
\partial r^2} \left\{ 
\int_0^{\infty} d\xi\, \xi\, e^{-\xi^2}
\left( {2 r\xi \over 3} - 2 \sqrt{\xi \over 3} {\sinh(r \sqrt{3 \xi})-
\sin(r \sqrt{3 \xi})\over \cosh(r \sqrt{3 \xi})-\cos(r \sqrt{3 \xi})}\right)
\right\} \cr & = {12^n \over 2 \sqrt{\pi} n^{7/4}} \times 3 \Phi'(r) = 
{12^n \over 2 \sqrt{\pi} n^{7/4}} \times 3 \rho(r) \ .}}
Up to a factor of $3$, this last expression is nothing but the number of 
quadrangulations with simply a marked vertex at distance $i=r\, n^{1/4}$ 
from a given origin. We therefore recover in the scaling limit the same 
ratio $3 \times 2^i$ as obtained in Eq.\avgeo\ for $1\ll i\ll n^{1/4}$
between geodesics of length $i$ and vertices at distance $i$. Up to this 
degeneracy factor, the density of geodesics of rescaled length $r$ 
coincides with the density of vertices at distance $r$. 

We can easily play the same game with $k$-tuples of weakly confluent 
geodesics. Using the expansion
\eqn\expanklimit{{1\over (2^i)^k} (Z_i^k-A_i^{(k)}) = {k \over 3^{k-1}}
{1\over n^{1/4}} \left({\cal F}(r,\xi) - {2\over 3\, r}\right)
+{\cal O}\left({1\over n^{1/2}}\right)}
we now get, for $t=e^{- s n^{-1/4}}/2^k$,
\eqn\laplaceUk{{\hat U}^{(k)}(t)=\hat \alpha^{(k)}(t) +
{k\, 3^{k+1} \over n^{1/2}} \int_0^{\infty} dr\, e^{-s\, r} 
\left({\cal F}''(r,\xi) -{4\over 3\, r^3}\right)  
+{\cal O}\left({1\over n^{3/4}}\right)
\ ,}
leading to
\eqn\scalUk{\left.{U^{(k)}_i\over (2^i)^k}\right\vert_{g^n} \buildrel 
{n\to \infty} \over \sim {12^n \over 2 \sqrt{\pi} n^{7/4}}\  k\, 3^k \rho(r) 
\ .}
Again, in the scaling limit, the density of $k$-tuples of weakly avoiding
confluent geodesics of length $r$ coincides with the density $\rho(r)$ 
upon a simple renormalization by a degeneracy factor $k (3 \times 2^i)^k$, 
in agreement with Eq.\avekgeo.

In the case $k=2$ and by a slight refinement of our generating functions,
we may compute the average number of contacts between two weakly avoiding
confluent geodesic of rescaled length $r$. This calculation is detailed
in Appendix A, with the result:
\eqn\contactr{\langle c \rangle_r = n^{1/4}\, {r\over 3}\ ,}
which extends the finite $i$ linear dependence displayed in Eq.\contact\ to 
the whole range of $r$ in the scaling regime.

\subsec{Exceptional points}
The situation is more interesting for exceptional points as it will lead 
us to a new scaling function. We consider again the case of two strongly
avoiding confluent geodesics of length $i$, as counted by $U_i^2$. 
{}From Eq.\scforU, we now get for $i=r\, n^{1/4}$ the scaling behavior
\eqn\scalUtwo{\eqalign{\left.{(U_i)^2\over (2^i)^2}\right\vert_{g^n} &
\buildrel {n\to \infty} \over \sim {12^n \over {\rm i} \pi n^{5/2}}
\int_{-\infty}^{+\infty} d\xi\, \xi\, e^{-\xi^2} \left(9 {\cal F}''(r, \xi) 
\right)^2 \cr &= 
{12^n \over 2 \sqrt{\pi} n^{5/2}} (2 \times 3^2) \ \sigma(r) \cr}}
where
\eqn\valsigma{\eqalign{& \sigma (r) \equiv {432 \over \sqrt{\pi}} \int_0^\infty
d\xi\, \xi^4\, e^{-\xi^2} \left\{8 {(c{\tilde c}-1)^3 \over (c-{\tilde c})^6}
+2{ (c{\tilde c}-1)(c{\tilde c}-4) \over (c-{\tilde c})^4} -{1\over 
(c-{\tilde c})^2}\right\} \cr 
&{\rm with} \ \ c\equiv \cosh (r \sqrt{3\xi})\  , \quad {\tilde c}
\equiv \cos(r \sqrt{3 \xi})\ .\cr} }
\fig{The scaling function $\sigma(r)$ counting the number of exceptional
points at a rescaled distance $r$ form the origin in the scaling limit
of large quadrangulations. There are on average $n^{1/4} \sigma(r) dr$ such
exceptional points at a rescaled distance between $r$ 
and $r+dr$. }{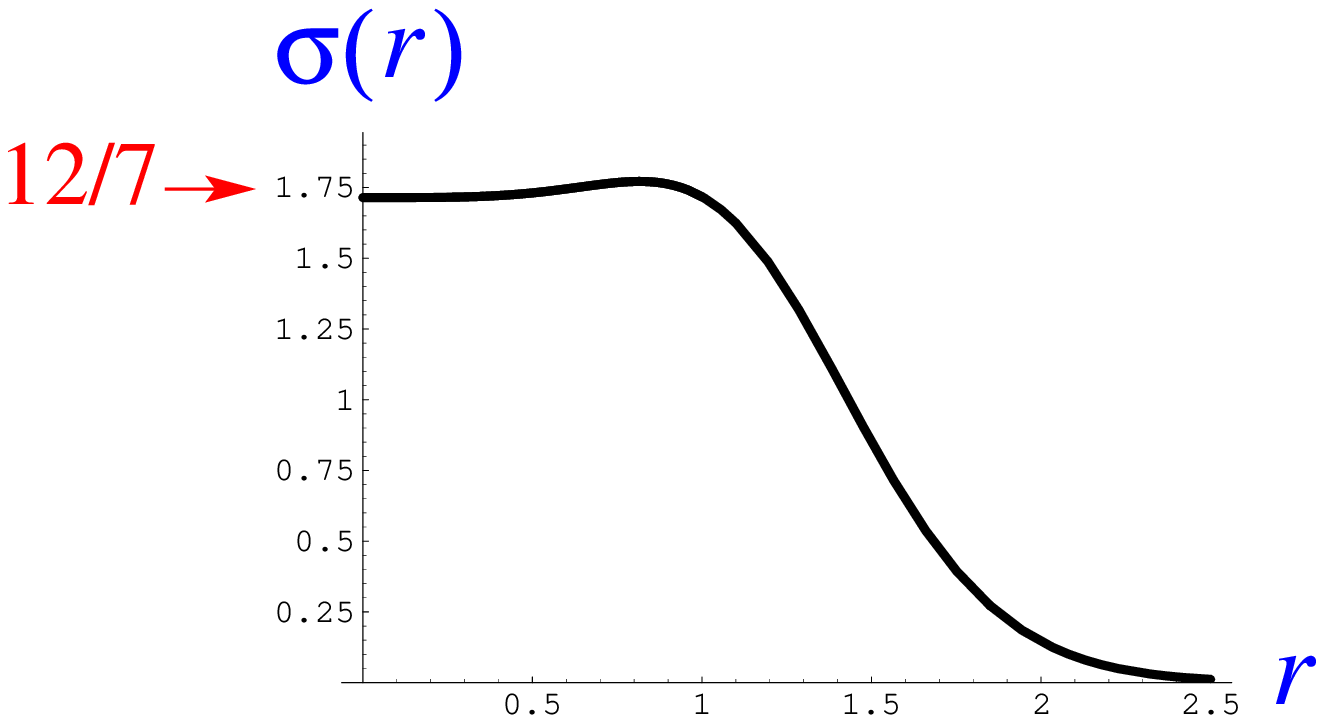}{10.cm}
\figlabel\exceptional
We have factored out in Eq.\scalUtwo\ the asymptotic number $12^n/(2 \sqrt{\pi}
n^{5/2})$ of quadrangulations with a given origin as well as a degeneracy 
factor $2 \times 3^2$ for two confluent geodesics. We may then interpret
the remaining finite scaling function $\sigma(r)$ as counting the number of 
exceptional points at a rescaled distance $r$ from a given origin. 
This function is plotted in Fig.\exceptional\ and has the following 
limiting behaviors:
\eqn\limbeha{\eqalign{& \sigma(r) \buildrel {r\to 0} \over
\rightarrow 12/7 \cr
&\sigma(r) \buildrel {r\to \infty} \over \sim e^{-3 (3/2)^{2/3} r^{4/3}}
\ .\cr}}
Note that the small $r$ limit corroborates the finite $i$ result of 
Section 3.4. Finally, upon summing over $i$, the total number of exceptional 
points is simply given by  $n^{1/4} \int_0^\infty dr \sigma(r)$. 

\subsec{Area between two geodesics}
Let us finally address again the question of the dispatching of the area
between geodesics, now in the scaling limit. Considering two non-intersecting 
geodesics, we shall call $\omega_1$ and $\omega_2$ the fraction of the total 
area $n$ carried by the two domains which they separate, with of course
$\omega_1+\omega_2=1$. 
A convenient way to access the average properties of $\omega_1$ and
$\omega_2$ is to introduce extra weight parameters $\lambda_1$ and 
$\lambda_2$ coupled to the area of the two domains. In the case of 
weakly avoiding geodesics, we therefore consider the refined generating 
function $Z_i(\lambda_1 g)Z_i(\lambda_2 g)$ counting quadrangulations with a
weight $\lambda_1 g$ (resp. $\lambda_2 g$) per square in the first (resp. second)
domain. Now we may explore the scaling limit of 
$Z_i(\lambda_1 g)Z_i(\lambda_2 g)$ by taking $\lambda_i=1+{\cal O}(1/n)$, 
namely by writing
\eqn\scallambda{\eqalign{ 
& g={1\over 12} \left(1+{\xi^2\over n}\right) \cr
& \lambda_1 g={1\over 12} \left(1+{\mu_1^2\over n}\right) +{\cal O}
\left({1\over n^2}\right) \cr
& \lambda_2 g={1\over 12} \left(1+{\mu_2^2\over n}\right) +{\cal O}
\left({1\over n^2}\right) \ , \cr}}
with $\mu_i\equiv \mu_i(\lambda_i,\xi)$ such that $\mu_i(1,\xi)=\xi$. 
At large $n$ and for $i=r\, n^{1/4}$, we have the expansion:
\eqn\zoneztwo{{1\over (2^i)^2} Z_i(\lambda_1 g) Z_i(\lambda_2 g)
=  {1\over 3^2} +{{\cal F}(r,\mu_1)+{\cal F}(r,\mu_2) \over
3 n^{1/4}} +{\cal O}\left({1\over n^{1/2}}\right) \ .}
Taking the corresponding irreducible part leads to a generating function 
$U^{(2)}_i(g;\lambda_1,\lambda_2)$ with expansion
\eqn\Uonetwo{{1\over (2^i)^2} U_i^{(2)}(g;\lambda_1,\lambda_2)
= {27 \left({\cal F}''(r,\mu_1)+{\cal F}''(r,\mu_2) \right)
\over n^{3/4}} +{\cal O}\left({1\over n}\right)}
and we end up with 
\eqn\Utwo{\left.{U_i^{(2)}(g;\lambda_1,\lambda_2) \over (2^i)^2}
\right\vert_{g^n} 
\buildrel {n\to \infty} \over \sim {12^n \over \pi n^{7/4}}\ 27 
\int_{-\infty}^{+\infty} d\xi\, \xi\, e^{-\xi^2} 
\left\{ {\cal F}''(r,\mu_1)+{\cal F}''(r,\mu_2)\right\}\ .}
To extract $\langle \omega_i\rangle $ ($i=1,2$), we must apply the operator
$(\lambda_i/n) (\partial/\partial \lambda_i)$ and let $\lambda_i\to 1$, 
i.e.\ $\mu_i\to \xi$.
At leading order in $n$, we have $(\lambda_i/n) (\partial/\partial \lambda_i)
=1/(2\mu_i) (\partial/\partial \mu_i)$ and we thus get
\eqn\aveomega{\langle \omega_1\rangle_r = \langle \omega_2\rangle_r
= {\int_{-\infty}^{+\infty} d\xi\, \xi\, e^{-\xi^2} {1\over 2 \xi}
{\partial \over \partial \xi} {\cal F}''(r, \xi) \over  
\int_{-\infty}^{+\infty} d\xi\, \xi\, e^{-\xi^2} 2 {\cal F}''(r, \xi)}
={1\over 2} }
as seen by an integration by part. This average result simply expresses 
the symmetry between the two domains and does not tell us anything about 
the actual dispatching of the area. As before, this dispatching is better 
measured by the correlation $\langle \omega_1 \omega_2 \rangle$. 
This latter average requires the action of the operator 
$1/(4 \mu_1 \mu_2) \partial^2/\partial \mu_1\partial \mu_2$
on the integrand in \Utwo, which produces a vanishing result, namely
\eqn\omom{ \langle \omega_1 \omega_2 \rangle_r= 0 }
in the limit $n\to \infty$. In other word, the fraction of the area of
the smaller domain tends to $0$ at large $n$. A closer look at the 
corrections to scaling leads to  $\langle \omega_1 \omega_2 \rangle_r 
= {\cal O}(1/n^{1/4})$, with an $r$-dependent coefficient. An integral
formula for this coefficient is presented in Appendix B. In particular, 
for small $r$, it behaves as $27 r^3 /100$, which is precisely 
the result announced in Eq.\lartheta\ when $i=r\, n^{1/4}$.
\fig{The scaling function $\lambda(r)$ of Eq.(4.31) and the ratio 
$\lambda(r)/\sigma(r)$ measuring the correlation $\langle \omega_1
\omega_2\rangle_r$ for the fractions of area $\omega_1$ and $\omega_2$
of the two domains delimited by two strongly avoiding confluent geodesics.
The average is for geodesics leading to a given exceptional point at 
distance $r$ from the origin. For small $r$, $\lambda(r)/\sigma(r) 
\sim (3/20)\, r^4$, which means that the larger fraction of area is 
of order $1$ and the smaller of order $r^4$. For large $r$, 
$\lambda(r)/\sigma(r) \to 1/4$, which means that both fractions tend
to $1/2$. 
}{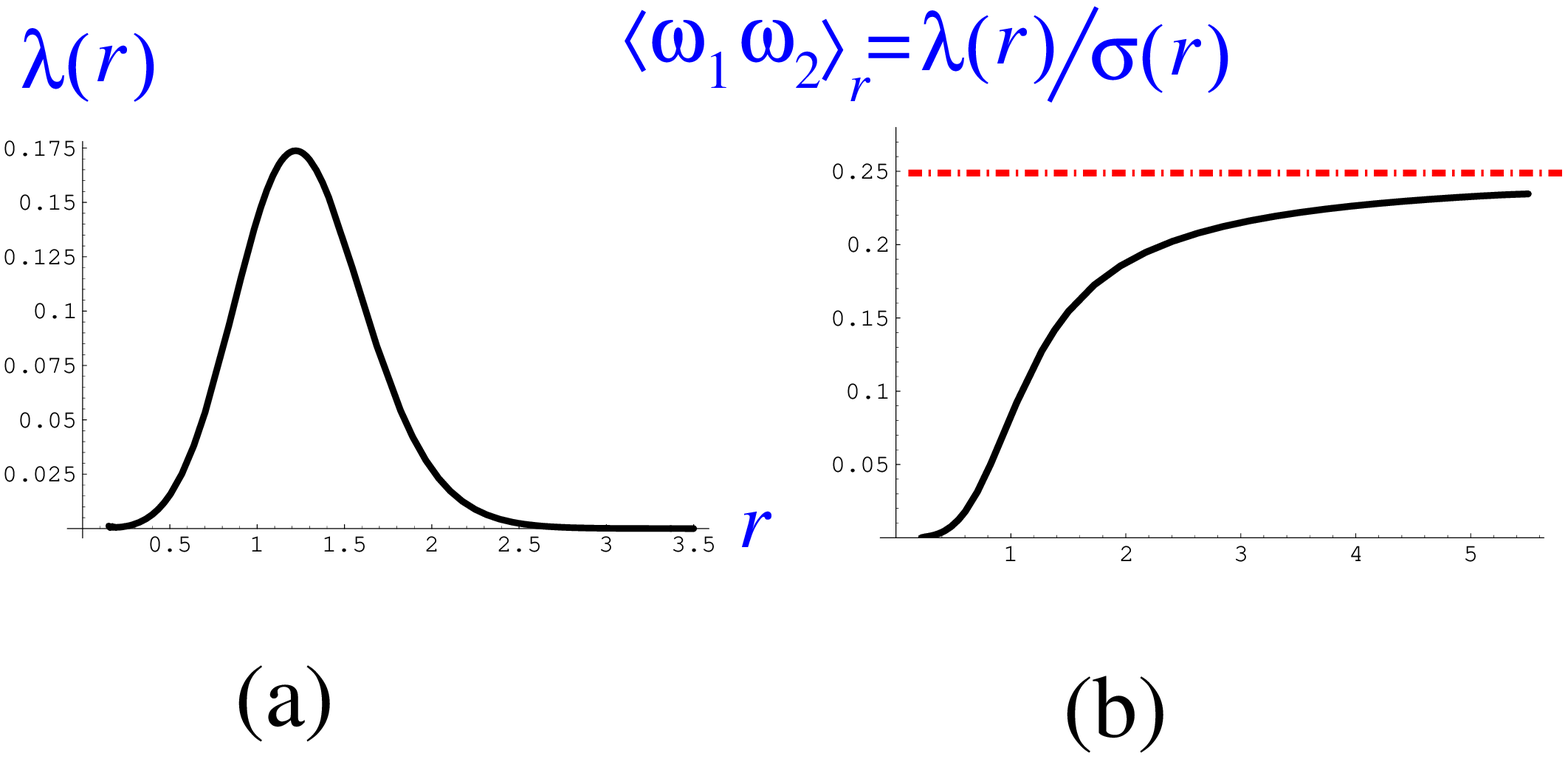}{13.cm}
\figlabel\lambdasigma
This situation is to be contrasted with the case of two strongly
avoiding geodesics, for which we must consider directly the expansion of 
$U_i(\lambda_1 g) U_i(\lambda_2 g)$, namely
\eqn\Uonetwobis{\left.{1\over (2^i)^2} U_i(\lambda_1 g) U_i(\lambda_2 g)
\right\vert_{g^n}=  {12^n \over {\rm i} \pi n^{5/2}}
\int_{-\infty}^{+\infty} d\xi\, \xi \, 
e^{-\xi^2} 81 {\cal F}''(r,\mu_1){\cal F}''(r,\mu_2)\ . }
Applying again $1/(2 \mu_i) \partial/\partial\mu_i$, we recover
$\langle \omega_1 \rangle_r= \langle \omega_2 \rangle_r =1/2$, while
we now obtain
\eqn\excepomega{\langle \omega_1\omega_2 \rangle_r = 
{\int_{-\infty}^{+\infty} d\xi\, \xi \,
e^{-\xi^2} \left({1\over 2 \xi }{\partial{\cal F}''\over \partial \xi} 
(r,\xi)\right)^2
\over \int_{-\infty}^{+\infty} d\xi\, \xi \,
e^{-\xi^2} \left({\cal F}''(r,\xi)\right)^2
}= {\lambda(r) \over \sigma(r)}}
where
\eqn\vallamb{\eqalign{\lambda(r)={1\over \sqrt{\pi} }
\int_0^{\infty} d\xi\, e^{-\xi^2}&\Big\{ 81 \left(3+2 \xi 
{\partial\over \partial \xi}\right) \left(
8 {(c{\tilde c}-1)^3 \over (c-{\tilde c})^6}
+2{ (c{\tilde c}-1)(c{\tilde c}-4) \over (c-{\tilde c})^4} -{1\over 
(c-{\tilde c})^2}\right) \cr & + 324 \xi r^2 s\, {\tilde s}
\left(36 {(c{\tilde c}-1)^3 
\over (c-{\tilde c})^8}+ 6 { (c{\tilde c}-1)(2c{\tilde c}-5) \over 
(c-{\tilde c})^6}+ {c{\tilde c}-4 \over (c-{\tilde c})^4}
\right)\Big\}\cr}}
with $c\equiv \cosh (r \sqrt{3\xi})$, ${\tilde c} \equiv \cos(r \sqrt{3 \xi})$,
$s\equiv \sinh (r \sqrt{3\xi})$ and ${\tilde s} \equiv \sin(r \sqrt{3 \xi})$.
This function is plotted in Fig.\lambdasigma\ together with the average
value $\langle \omega_1 \omega_2\rangle_r =\lambda(r)/\sigma(r)$.
At small $r$, the quantity between curly brackets in \vallamb\ behaves 
as $(36/35) \xi^2 r^4$, leading 
to $\lambda(r) \sim (9/35)\, r^4$ and $\langle \omega_1\omega_2 \rangle_r
\sim (3/20)\, r^4$. We thus recover an area of order $n$ 
in the larger domain and an area of order $(3/20)\, n\, r^4= (3/20)\, i^4$
in the smaller one, in agreement with the finite $i$ result \masmas.
For an arbitrary finite value of $r$, both $\omega_1$ and $\omega_2$ are 
finite, leading to two domains whose area is proportional to $n$.
This indicates that exceptional points are reachable by two geodesics
which are truly distinct, even in the continuum limit, as they separate
macroscopic domains of extensive area. Finally, for large $r$, $\lambda(r)$
vanishes like $e^{-3 (3/2)^{2/3} r^{4/3}}$ as $\sigma(r)$, with 
a limiting ratio
\eqn\limrat{{\lambda(r)\over \sigma(r)}\buildrel {r\to \infty} \over
\rightarrow {1\over 4}\ .}
Now we note that we have necessarily $\omega_1 \omega_2 \leq 1/4$ since
$\omega_1$ and $\omega_2$ are real positive numbers adding to $1$ and
we have equality only for $\omega_1=\omega_2=1/2$. The large $r$
limiting average value $\langle \omega_1 \omega_2 \rangle=1/4$ therefore
indicates that the exceptional points sitting at
a large rescaled distance $r$ from the origin separate the quadrangulation 
in two domains of equal area $n/2$. Note that such exceptional points
may appear only in pointed quadrangulations with a rescaled radius larger
than $r$, which are exponentially rare.

\newsec{Conclusion}
In this paper, we have presented a number of exact results on the
statistics of geodesics in quadrangulations of large size $n$. 
At a discrete level, the number of geodesics between any two vertices at a 
distance $i$ from each other grows like $3\times 2^i$ at large $i$. 
For a generic pair of vertices, any two geodesics linking these vertices 
stick to each other in the sense that their number of contacts is proportional 
to $i$. Moreover, they separate the quadrangulation in two very asymmetric
domains, with the larger domain containing most of the area of the 
quadrangulation while the smaller one has an area of order $i^3 \ll n$. 
This symmetry breaking extends to the case of $k$-tuples of geodesics, for 
which only one of the $k$ domains which they form in the quadrangulation has
an area of order $n$. 

Upon factoring out the entropy factor $3\times 2^i$, a sensible scaling 
limit can be reached in the regime $i= r n^{1/4}$. In this limit, the density 
of geodesics of rescaled length $r$ simply coincides with the density of 
vertices at distance $r$. Similarly, for $k$-tuples of geodesics, we 
obtain the same density upon dividing by the entropy factor 
$k\, (3\times 2^i)^k$, where the first $k$ prefactor is an expression
of the above-mentioned symmetry breaking phenomenon. A possible interpretation
of these results is that there is essentially a unique geodesic path 
between any two points in the scaling limit and that the entropy factors
simply account for a degeneracy at a microscopic level. More studies are
however necessary to validate this picture. In particular, we noted that 
the area enclosed by two geodesics, although negligible with
respect to $n$, still scales as $n^{3/4}$, i.e.\ faster than
the length $i \propto n^{1/4}$ of the geodesics. Viewing the two geodesics
as forming a sequence of blobs without contact, this means that at least
one of these blobs has an area larger than $n^{1/2}$. It is 
not clear whether the observed scaling is sufficient to guarantee that 
we can simply ignore the domain trapped between the geodesics in the scaling
limit.

In addition to this generic behavior, we also found that, for a given
origin vertex, there is a number of order $n^{1/4}$ of exceptional points
that can be linked to this origin by two truly distinct geodesics with
no contacts on the way. Both domains separated by the geodesics have an
area proportional to $n$ in this case. We explicited the density of these 
exceptional endpoints in the scaling limit as a function of the rescaled 
distance $r$ but a more detailed analysis of their correlations would
be desirable to understand better the geometry of this set of points. 

It would be nice to extend our results to other families of maps.
We expect in general that the precise value of the degeneracy factors
counting discrete geodesic paths will depend on the family at hand but
that all our results in the scaling limit should remain unchanged.
This holds for maps in the universality class of so-called
``pure gravity". Studying models of maps with matter degrees of freedom is another 
story which requires new insights.

\bigskip
\noindent{\bf Acknowledgments:}
We thank F. David, P. Di Francesco and J.-F. Le Gall for helpful discussions.
The authors acknowledge support from the Geocomp project, ACI Masse de 
donn\'ees, from the ENRAGE European network, MRTN-CT-2004-5616 and from 
the Programme d'Action Int\'egr\'ee J. Verne ``Physical applications of 
random graph theory".

\appendix{A}{Average number of contacts between two confluent geodesics}

We compute here in the scaling limit $i=r\, n^{1/4}$ the average number
of contact vertices between two weakly avoiding confluent geodesics 
of length $i$. To this end, it is convenient to introduce the generating
function $Z_i^{pp}(g)$ of quadrangulations with a geodesic boundary of 
length $i$ and with a {\it marked pinch point}. 
We have the relation:
\eqn\zpp{Z_i^{pp}=\sum_{j=1}^{i-1} Z_j\, Z_{i-j}\ ,}
which simply expresses that the marked pinch point, at arbitrary distance
$j$, separates the quadrangulation with a boundary in two parts of 
length $j$ and $i-j$. 

To treat the case of
two geodesic paths with a marked contact, we simply have to consider
the generating function $2 Z_i Z_i^{pp}$, which counts quadrangulations with
a geodesic boundary of length $2i$, with a marked geodesic path in-between, 
and with a marked contact between this geodesic path and either side of 
the boundary. Note that if the two boundaries come into contact, the 
corresponding pinch point is actually counted twice. For a proper
enumeration of quadrangulations with marked geodesics, we must as usual 
take an irreducible part. In the same way as the irreducible part 
$U_i^{(2)}$ of $Z_i^2$ counts quadrangulations with two weakly avoiding 
confluent geodesics, the irreducible part $U_i^{(2)pp}$ of $2 Z_i Z_i^{pp}$ 
counts these same objects with a marked contact. 
It is obtained by removing configurations where the two boundaries come into contact,
which translates into the relation
\eqn\uipp{U_i^{(2)pp}= (2 Z_i Z_i^{pp}) - \sum_{j=1}^{i-1} \left( U^{(2)}_j
\times 2 Z_{i-j} Z^{pp}_{i-j}+ U^{(2)}_j\times 2 Z_{i-j}^2+ U^{(2)pp}_j \times
Z_{i-j}^2\right)\ ,}
where the subtracted terms correspond to configurations where the two boundaries 
have a first contact at a distance $j$ from the origin. This contact 
may be before the marked pinch point (first term in the sum), 
at the marked pinch point 
(second term) or after the marked pinch point (third term). 

As before, the above relations transform into simple relations for the 
generating functions 
\eqn\laplap{\eqalign{
&\hat Z(t)\equiv \sum_{i=1}^\infty Z_i t^i 
\ , \quad \hat Z^{(2)}(t) \equiv \sum_{i=1}^\infty Z_i^2 t^i 
\ , \quad \hat Z^{pp}(t)\equiv \sum_{i=1}^\infty Z_i^{pp} t^i \cr
&\hat Z^{(2)pp}(t)\equiv \sum_{i=1}^\infty 2 Z_i Z_i^{pp} t^i 
\ , \quad \hat U^{(2)pp} (t) \equiv \sum_{i=1}^\infty U_i^{(2)pp} t^i
\ , \quad \hat U^{(2)}(t)\equiv \sum_{i=1}^\infty U_i^{(2)} t^i \ .\cr}}
Indeed, Eq.\zpp\ translates into 
\eqn\ZtoZZ{\hat Z^{pp} (t)= \left(\hat Z(t)\right)^2}
and, using the relation $\hat U^{(2)}(t)= \hat Z^{(2)}(t) /\left
(1+\hat Z^{(2)}(t)\right)$, Eq.\uipp\ becomes simply
\eqn\ZtoUpp{\hat U^{(2)pp}(t)= {\hat Z^{(2)pp}(t) - 2 \left(\hat Z^{(2)}(t)
\right)^2 \over \left(1+ \hat Z^{(2)}(t)\right)^2 }\ .}

As in Section 4.2, in order to expand all the above generating functions
in the scaling limit, it is convenient to treat separately their values 
right at $\xi=0$, i.e.\ when $g=g_{\rm crit}=1/12$. Beyond the quantities 
$A_i\equiv Z_i(1/12)$, $A_i^{(2)}\equiv (A_i)^2$ and their generating 
functions $\hat A(t)$ and $\hat A^{(2)}(t)$ already defined in Sections 3.2 
and 3.3, we also introduce the notations
\eqn\moreAs{A^{pp}_i \equiv \sum_{j=1}^{i-1} A_j A_{i-j}\ , \quad
\hat A^{pp}(t) \equiv \sum_{i-1}^\infty A_i^{pp} t^i = \left(\hat A(t)
\right)^2\ , \quad \hat A^{(2)pp}(t)=\sum_{i-1}^\infty 2 A_i A_i^{pp} t^i \ .}
{}From Eqs.\ZtoZZ, \laplaceZ\ and \exat, we have the expansion
\eqn\laplaceZtwo{
\hat Z^{pp}\left({1\over 2}\, e^{-s n^{-1/4}} \right)
-\hat A^{pp}\left({1\over 2}\, e^{-s n^{-1/4}} \right)
= {2 n^{1/4} \over 3 s} \int_0^{\infty} 
dr\, e^{-s\, r} \left({\cal F}(r,\xi)-{2\over 3\, r}\right)
+\ldots }
from which we deduce the scaling behavior
\eqn\expanzpp{{1\over 2^i}(Z_i^{pp}-A_i^{pp}) = {2 \over 3}
\int_0^r dr'\, \left({\cal F}(r',\xi)-{2\over 3\, r'}\right) +\ldots }
This expansion, together with the expansion \elevexp\ for $Z_i-A_i$ leads
to 
\eqn\expanzppbis{{1\over (2^i)^2}( 2 Z_i Z_i^{pp}-2 A_i A_i^{pp}) = 
{2 r \over 9} \left({\cal F}(r,\xi)-{2\over 3\, r}\right)+ {4 \over 9}
\int_0^r dr'\, \left({\cal F}(r',\xi)-{2\over 3\, r'}\right) +\ldots }
where we also made use of the large $i$ leading values $A_i/2^i\sim 1/3$ and 
$A_i^{pp}/2^i \sim i/9= n^{1/4} r/9$. 
This yields 
\eqn\laplZtwopp{\eqalign{
\hat Z^{(2)pp}\left({1\over 4}\, e^{-s n^{-1/4}}\right)&
-\hat A^{(2)pp}\left({1\over 4}\, e^{-s n^{-1/4}}\right) \cr
= {2 n^{1/4} \over 9 } \int_0^{\infty} 
dr\, e^{-s\, r} & \left\{ r\, \left({\cal F}(r,\xi)-{2\over 3\, r}\right)+ 2 
\int_0^r dr' \left({\cal F}(r',\xi)-{2\over 3\, r'}\right)
\right\}
+\ldots\cr} }
while, from
\eqn\expztwo{{1\over (2^i)^2} (Z_i^2-A_i^2) = 
{2 \over 3 n^{1/4}} \left({\cal F}(r,\xi)-{2\over 3\, r}\right)+ \ldots }
we have
\eqn\lapZtwo{
\hat Z^{(2)}\left({1\over 4}\, e^{-s n^{-1/4}}\right)-
\hat A^{(2)}\left({1\over 4}\, e^{-s n^{-1/4}}\right)
={2 \over 3 } \int_0^{\infty} 
dr\, e^{-s\, r} \left({\cal F}(r,\xi)-{2\over 3\, r}\right) +\ldots }
Inserting the expansions \laplZtwopp\ and \lapZtwo\ in \ZtoUpp\ and using
the leading behaviors $\hat A^{(2)}\left(e^{-s n^{-1/4}}/4\right) \sim
n^{1/4}/(9s)$ and $\hat A^{(2)pp}\left(e^{-s n^{-1/4}}/4\right) \sim
2 n^{1/2}/(27s^2)$, we obtain finally
\eqn\lapures{\eqalign{ 
\hat U^{(2)pp}\left({1\over 4} e^{-s n^{-1/4}}\right)& 
-\hat W^{(2)pp}\left({1\over 4} e^{-s n^{-1/4}}\right) \cr
& ={1\over n^{1/4}}\int_0^\infty dr\, e^{-s\, r} \Big\{
-72 s \left({\cal F}(r,\xi)-{2\over 3\, r}\right) \cr & \ \  
+18 s^2 r\, \left({\cal F}(r,\xi)-{2\over 3\, r}\right)+ 36 s^2 \int_0^r dr' 
\left({\cal F}(r',\xi)-{2\over 3\, r'}\right) \Big\}+\ldots \cr 
&={1\over n^{1/4}}\int_0^\infty dr\, e^{-s\, r} 18 r\, 
\left({\cal F}''(r,\xi)-{4\over 3\, r^3}\right) +\ldots\cr} }
upon integrations by part over $r$. Here we used the notation
\eqn\genw{\hat W^{(2)pp}(t) \equiv {\hat A^{(2)pp}(t)-2\left( \hat A^{(2)}(t)
\right)^2\over \left( 1+\hat A^{(2)}(t)\right)^2}}
which is the critical value of $U^{(2)pp}(t)$ at $g=1/12$. 
We end up with
\eqn\expanuipp{{1\over (2^i)^2} (U_i^{(2)pp}-W_i^{(2)pp}) =
{1 \over n^{1/2}}\left(18\, r\, {\cal F}''(r,\xi)-{24\over r^2}\right)+\ldots}
where $W_i^{(2)pp}$ is the coefficient of $t^i$ in $\hat W^{(2)pp}(t)$.
{}From the exact expression for $A_i$, we can easily compute $\hat W^{(2)pp}(t)$
exactly. In particular, we have the small $\eta$ expansion:
\eqn\expw{\hat W^{(2)pp}\left({1\over 4}(1-\eta)\right)=
4+(12-8 \pi^2 +24 \log(\eta))\eta +{\cal O}(\eta^2)\ .}
{}From the singularity at $\eta=0$, we deduce that $W_i^{(2)pp}/(2^i)^2\sim
24/i^2 = 24/(n^{1/2}r^2)$ in the scaling regime, and therefore
\eqn\expanuippfin{{1\over (2^i)^2} U_i^{(2)pp}=
{18 \over n^{1/2}}\, r\, {\cal F}''(r,\xi)+\ldots}
Upon integration over $\xi$, this gives finally
\eqn\scalfin{\left.{U_i^{(2)pp}\over (2^i)^2}\right\vert_{g^n}
\buildrel {n\to \infty} \over \sim {12^n \over \pi n^{3/2}}\, 18\, r  \, 
\int_{-\infty}^{+\infty} d\xi\, \xi\, e^{-\xi^2} {\cal F}''(r,\xi)\ .}
Dividing this result by 
\eqn\scalfintwo{\left.{U_i^{(2)}\over (2^i)^2}\right\vert_{g^n}
\buildrel {n\to \infty} \over \sim {12^n \over \pi n^{7/4}}\, 54  \, 
\int_{-\infty}^{+\infty} d\xi\, \xi\, e^{-\xi^2} {\cal F}''(r,\xi)}
yields the average number of contacts 
\eqn\avecon{\langle c \rangle_r = n^{1/4}\, {r \over 3}\ .}
For small $r$, this gives the finite $i$ result $\langle c \rangle_i 
= {i \over 3}$ of Eq.\contact. The linear dependence of the average number
of contacts extends in practice to the whole range of $r$ in the scaling limit.

\appendix{B}{Correction to the area correlation for two weakly avoiding 
geodesics}

The scaling result \omom, i.e.\ $\langle \omega_1 \omega_2\rangle_r=0$ for 
two weakly avoiding confluent geodesics holds in the scaling limit 
$n\to \infty$. We can easily get the first correction to scaling by
computing the $1/n$ term in Eq.\Uonetwo\ giving 
$U_i^{(2)}(g;\lambda_1,\lambda_2)/(2^i)^2$. In practice, as we eventually
apply the operator $1/(4 \mu_1 \mu_2) \partial^2/\partial \mu_1\partial \mu_2$,
we need only the corresponding cross term depending on both $\mu_1$ and
$\mu_2$. This term reads
\eqn\crossterm{\eqalign{& {81\over n} \Big\{ {\partial^2\over \partial r^2} 
\left({\cal F}(r,\mu_1) {\cal F}(r,\mu_2)-{4\over 9\, r^2} \right) \cr
\ \ \ \ \ \ \ \ \ \ \ \ \ \ \ &-2 
{\partial^3\over \partial r^3} \left(\int_0^r dr'\, \left({\cal F}(r',\mu_1)
-{2\over 3\, r'}\right)\left(
{\cal F}(r-r', \mu_2)-{2\over 3\, (r-r')}\right)\right)\Big\}\cr}}
This leads immediately to the correction
\eqn\omomcorrec{ \eqalign{ \langle \omega_1 \omega_2\rangle_r= 
{9\over n^{1/4} \rho(r)}{1\over {\rm i}\sqrt{\pi}} &
\int_{-\infty}^{+\infty} d\xi\,\xi\,e^{-\xi^2}\Big\{
{\partial^2 \over \partial r^2} \left({1\over 2 \xi} {\partial{\cal F}
\over \partial \xi}(r, \xi) \right)^2\cr & -2 {\partial^3 \over \partial r^3}
\left(\int_0^r dr' {1\over 2\xi}{\partial {\cal F}\over \partial \xi}(r',\xi) 
{1\over 2\xi}{\partial {\cal F}\over \partial \xi}(r-r',\xi)\right) \Big\}\cr}.}
In particular, at small $r$, we have the expansion
\eqn\smallr{{1\over 2 \xi}{\partial {\cal F}\over \partial \xi}
(r, \xi) = {r^3 \over 30} + {\rm i} \xi {r^5 \over 140} +\ldots}
so that the $\{\cdot \}$ term in the integrand of Eq. \omomcorrec\ behaves as
\eqn\integran{{3 \over 100} r^4 +{\rm i} \xi {9\over 350} r^6   + \ldots}
Upon integration over $\xi$, only the second term survives so that the
integral behaves as ${\rm i} \sqrt{\pi} \times 9 r^6/700$. From the
 small $r$ behavior $\rho(r) \sim 3r^3/7$, we deduce
\eqn\omomsmallr{\langle \omega_1 \omega_2\rangle_r \buildrel
{r\to 0} \over \sim {1\over n^{1/4}} {27\over 100} r^3}
which is nothing but the announced result \lartheta. 
\listrefs
\end